\documentclass[pra,superscriptaddress,showpacs,longbibliography,reprint]{revtex4-2} 

\usepackage[colorlinks=true,linkcolor=blue,urlcolor=blue,citecolor=blue,pdfusetitle]{hyperref}
\usepackage{color}
\usepackage[utf8]{inputenc}
\usepackage[usenames,dvipsnames]{xcolor}
\usepackage{amsthm}
\usepackage{amsmath}
\usepackage{amssymb}
\usepackage{graphicx}
\usepackage{bm}
\usepackage[all]{xy}
\usepackage{lipsum}
\usepackage{braket}
\usepackage{dsfont}
\usepackage{array}
\usepackage{makecell}
\usepackage{xr}
\usepackage{float}
\usepackage{simpler-wick}

\begin{document}
\raggedbottom

\title{Unraveling-induced  entanglement phase transition  in  diffusive trajectories of continuously monitored noninteracting fermionic systems}

\author{Moritz Eissler}
\affiliation{Institut f\"ur Theoretische Physik and Center for Integrated Quantum Science and Technology, Universit\"at T\"ubingen, Auf der Morgenstelle 14, 72076 T\"ubingen, Germany}

\author{Igor Lesanovsky}
\affiliation{Institut f\"ur Theoretische Physik and Center for Integrated Quantum Science and Technology, Universit\"at T\"ubingen, Auf der Morgenstelle 14, 72076 T\"ubingen, Germany}
\affiliation{School of Physics and Astronomy and Centre for the Mathematics and Theoretical Physics of Quantum Non-Equilibrium Systems, The University of Nottingham, Nottingham, NG7 2RD, United Kingdom}
\author{Federico Carollo}
\affiliation{Centre for Fluid and Complex Systems, Coventry University, Coventry, CV1 2TT, United Kingdom}

\begin{abstract}
    The competition between unitary quantum dynamics and dissipative stochastic effects, as emerging from continuous-monitoring processes,  can culminate in measurement-induced phase transitions. Here, a many-body system abruptly passes, when exceeding a critical measurement rate, from a highly entangled phase  to a low-entanglement one. We consider a different perspective on entanglement phase transitions  and explore whether these can emerge when the measurement process itself is modified, while keeping the measurement rate fixed. To illustrate this idea, we consider a noninteracting fermionic system and focus on diffusive detection processes. Through extensive numerical simulations, we show that, upon varying a suitable \textit{unraveling parameter} ---interpolating between measurements of different quadrature operators--- the system displays a transition from a phase with area-law entanglement to one where entanglement scales logarithmically with the system size. Our findings may be relevant for tailoring quantum correlations in noisy quantum  devices and for conceiving optimal classical simulation strategies.  
\end{abstract}

\maketitle

\section{Introduction} 
\begin{figure}[t]
\centering
\includegraphics[width=\columnwidth]{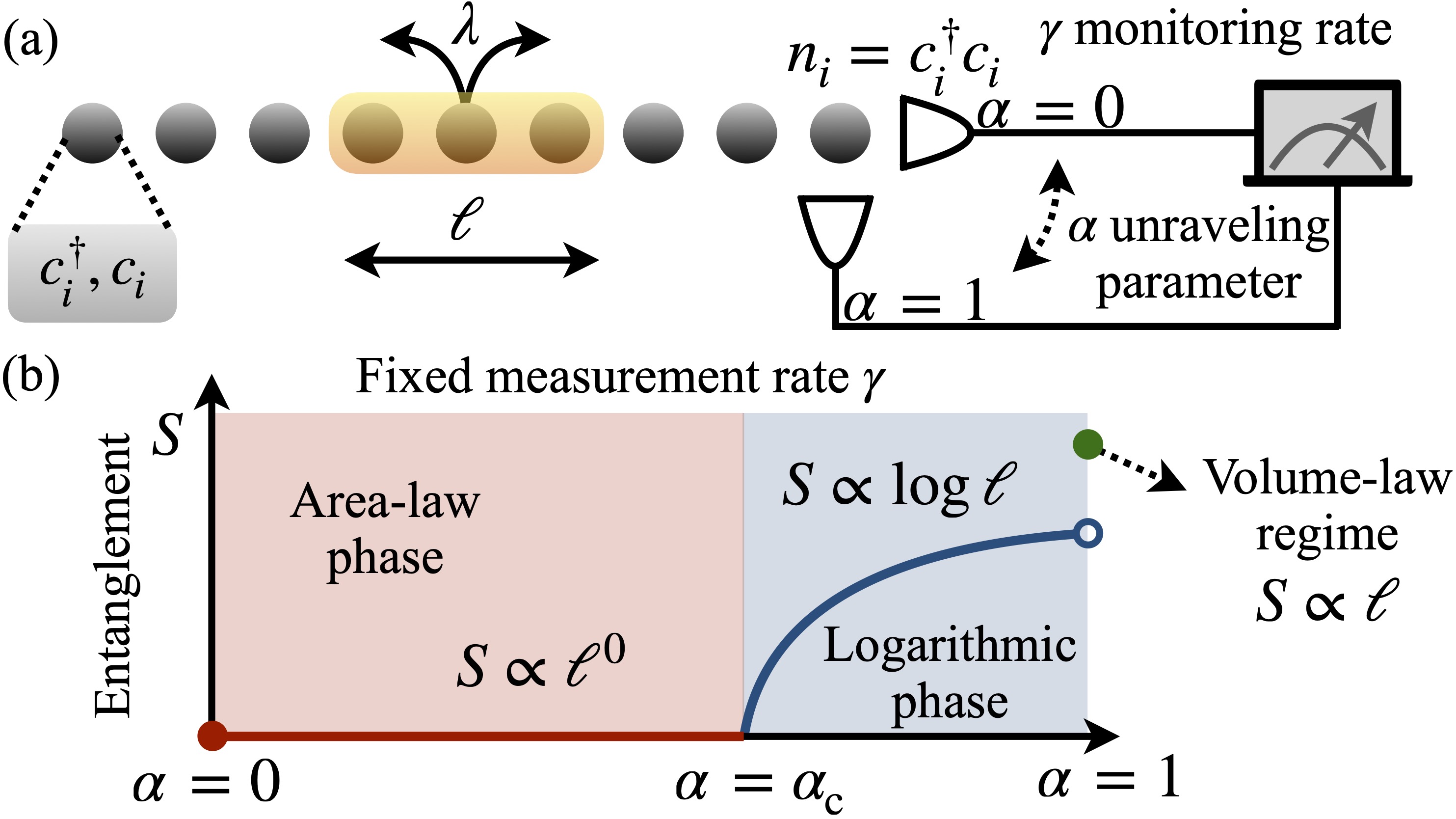}
      \caption{{\bf Setup and entanglement  transition.} (a) One-dimensional system with $L$ sites, occupied by fermionic particles (creation and annihilation operators $c_i^\dagger,c_i$). Fermions hop between neighboring sites with rate $\lambda$. The local excitation densities, $n_i$,  are continuously monitored, at a rate $\gamma$, through a detection scheme \cite{wiseman2001,wiseman2009,genoni2014,clarke2023}, which interpolates, via an \textit{unraveling parameter} $\alpha\in[0,1]$, between two different homodyne-detection measurements. The parameter $\alpha$ thus specifies the measurement process and  affects the stochastic dynamics of the system [cf.~Eq.~\eqref{Unravelling3}]. The sketch shows a bipartition of the system with a subsystem of length $\ell$.  (b) For $\alpha=0$, the system features area-law entanglement ($S\propto \ell^0$) while for $\alpha=1$  volume-law entanglement ($S\propto \ell$) is found.  At an intermediate value $\alpha=\alpha_{\rm c}$, the system displays a transition from unravelings with area-law ($\alpha<\alpha_{\rm c}$) to unravelings with subextensive logarithmic entanglement growth ($\alpha_{\rm c}<\alpha<1$).       }
      \label{setup}
  \end{figure}

Entanglement stands out as the most paradigmatic feature of quantum mechanics. Beyond its relevance in quantum information, the spreading of entanglement in many-body systems is tied to fundamental questions  \cite{10.21468/SciPostPhysLectNotes.20}, e.g., related to the emergence of critical correlations close to quantum phase transitions \cite{osterloh2002,Baggioli_2023,RevModPhys.80.517,PhysRevLett.90.227902} or of 
thermal ensembles in isolated quantum systems \cite{PhysRevE.50.888,Rigol_2008,PhysRevX.9.021027}. To shed light on these phenomena, basic   models of random quantum  circuits have been introduced \cite{PhysRevX.7.031016,Fisher_2023,Weinstein_2023,Cheng_2023,chen2023,PhysRevB.107.L140301}. Their analysis demonstrated that generic (nonintegrable) unitary systems evolve towards strongly correlated states, displaying volume-law entanglement \cite{Calabrese_2005}. That is, bipartite entanglement which grows  with the size $\ell$ of the smallest subsystem generated by a bipartition [see sketch in Fig.~\ref{setup}(a)].

Observing local properties of many-body quantum systems in real time, either through projective measurements 
\cite{PhysRevX.7.031016,PhysRevX.9.031009,PhysRevB.98.205136,PhysRevB.99.224307,PhysRevLett.125.030505,PhysRevB.101.104302,PhysRevB.101.235104,kelly2024} or through (weak) continuous-monitoring processes \cite{Jacobs_2006}, can dramatically affect the built-up of entanglement\cite{PhysRevB.98.205136,PhysRevX.9.031009,PhysRevB.100.134306,10.21468/SciPostPhys.7.2.024,PhysRevX.10.041020,Alberton_2021,Turkeshi_2021,Turkeshi_2022,Coppola_2022,PhysRevB.106.L220304,PhysRevLett.128.243601,Le_Gal_2023,PhysRevLett.130.230401,Ippoliti_2021,PhysRevResearch.2.013022,vovk2024quantum,PhysRevB.101.060301}. For a small measurement rate $\gamma$, the unitary volume-law behavior may be expected to survive the presence of a local monitoring. However, for a large rate $\gamma$, the quantum state necessarily remains close to a product state and thus features area-law entanglement, scaling with the size of the boundary of a bipartition [cf.~Fig.~\ref{setup}(a)]. Remarkably,  many-body systems can display  a genuine transition between these two phases, as a function of the measurement rate $\gamma$  \cite{PhysRevX.10.041020,PhysRevResearch.2.013022,PhysRevResearch.2.013022,PhysRevB.101.104301,PhysRevB.100.134306,PhysRevLett.125.070606,PhysRevResearch.2.023288,PhysRevB.101.060301}.  
For the case of noninteracting fermionic systems, measurement-induced transitions can occur from an area-law phase to one featuring entanglement growth that is logarithmic in $\ell$  \cite{Alberton_2021,minato2022}.

In this work, we take a different perspective on entanglement phase transitions. Rather than exploring the behavior of the system upon varying the measurement rate $\gamma$, we analyze the entanglement phases generated by varying the measurement process itself [see sketch in Fig.~\ref{setup}(a)]. The rationale is as follows. The dynamics of the quantum state averaged over all realizations of a continuous-monitoring process is described by a Markovian quantum master equation. Different monitoring processes, even when performed at the same rate $\gamma$, are associated with different ``unravelings" of the quantum master equation into quantum stochastic processes. It is thus important to understand how entanglement properties of  many-body systems depend on the considered measurement process, or unraveling. In Refs.~\cite{PhysRevLett.128.243601,vovk2024quantum}, this dependence was explored in the context of devising optimal simulation strategies, where an entanglement transition driven by a change of the unraveling dynamics was observed. However, the system considered there is a quantum circuit with a fully random Hamiltonian, for which one of the two unravelings of the measurement process can be re-absorbed in the Hamiltonian ensemble without modifying it. As such, the emergent transition can be exactly mapped on a measurement-induced one  since there is effectively a single relevant unravelling of the dissipative processes \cite{PhysRevLett.128.243601,vovk2024quantum}. The question thus remains whether an unraveling-induced phase transitions can occur for many-body systems possessing a deterministic Hamiltonian evolution and for which a mapping to a standard measurement-induced transition is not possible. 

To find an answer, we consider noninteracting fermionic systems and focus on two different unravelings, associated with homodyne-detection processes \cite{PhysRevLett.120.133601}. The first one  results in area-law entanglement while the second one in volume-law entanglement, as shown in Fig.~\ref{setup}(b). We then  construct a family of unravelings which interpolates between the  two processes \cite{wiseman2001,wiseman2009,genoni2014,clarke2023} and show that the system undergoes a transition from an area-law  to a logarithmic entanglement phase \cite{Alberton_2021,Coppola_2022}.

\section{Monitored tight-binding model}
Our study is based on a one-dimensional tight-binding lattice model, with periodic boundary conditions, which is a paradigmatic system for the study of transport and correlation spreading. The system, see Fig.~\ref{setup}(a), is made of $L$ sites, each one hosting a fermionic particle with the corresponding creation and annihilation operators $c_i^\dagger,c_i$. The Hamiltonian dynamics entails nearest-neighbor hopping represented by the operator 
\begin{align}
H=\sum_{i,j=1}^L h_{ij}c^{\dagger}_i c_j\, ,
\label{Hamiltonian}
\end{align}
with $h_{ij} = \lambda (\delta_{i,j+1}+\delta_{i,1}\delta_{j,L}+{\rm h.c.})$ and  $\lambda$ being the coherent hopping rate. As initial state, we take the N\'eel state 
\begin{align}
    \ket{\psi_0 }  = \prod_{j=1}^{L/2} c_{2j}^{\dagger} \ket{0}\,,
    \label{Neel-state}
\end{align}
with $\ket{0}$ being the fermionic vacuum, $c_j\ket{0}=0$, $\forall j$. 

When analyzing the dynamics of correlations, a relevant  quantity is the entanglement shared by a bipartition of the lattice. Here, we focus on the situation in which the system is partitioned into two equal halves, of length $\ell=L/2$. 
Under the  unitary dynamics governed by $H$, the wave function of the fermionic particles, initially localized on even sites [cf.~Eq.~\eqref{Neel-state}], spreads over the entire system entangling different parts of it. This propagation determines a linear growth  ($\sim\lambda t$) of entanglement, as quantified by the entanglement entropy of the bipartition. The latter is defined as $S=-{\rm Tr}_\ell \left(\rho^\ell \log \rho^\ell\right)$, where $\rho^\ell={\rm Tr}_{\ell}'\left(\ket{\psi}\!\!\bra{\psi}\right)$ and where ${\rm Tr}_\ell$, ${\rm Tr}_\ell'$ represent the trace over the sites in the subsystem and in the remainder of the system, respectively. When the particles are completely delocalized, the entanglement entropy saturates to a stationary volume-law value $S\sim \ell $ \cite{10.21468/SciPostPhys.7.2.024}. 

 \begin{figure}[t]      
        \includegraphics[width = \columnwidth]{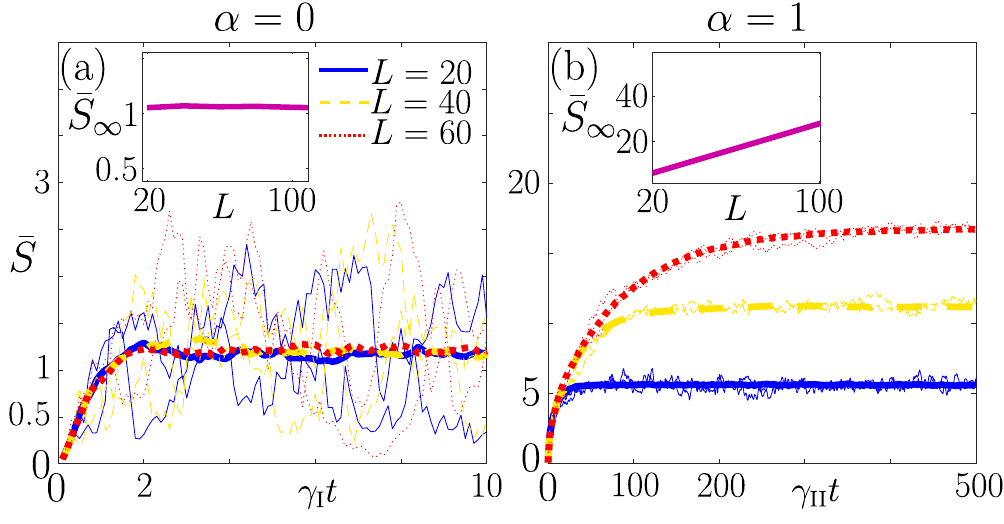}       \caption{\textbf{Entanglement entropy dynamics.} (a) Half-chain entanglement entropy, $\bar{S}_t$, averaged over $100$ realizations of the quantum process in Eq.~\eqref{Unravelling1} (bold lines) for different sizes $L=20,40,60$ (solid line being the smallest and dotted line being the largest size). Thin lines show  single realizations of the process. For the chosen value of the monitoring rate $\lambda=\gamma_{\rm I}/2$, the stationary average entanglement entropy $\bar{S}_\infty$ features area-law behavior as shown in the inset. 
        (b) Same as in (a) for the process in Eq.~\eqref{Unravelling2}, with $\lambda=\gamma_{\rm II}/2$. Entanglement here follows a volume-law behavior as shown in the inset.}
        \label{Fig2}
    \end{figure}

The emergence of volume-law entanglement may be hindered by local continuous-measurement processes. In a typical setting \cite{10.21468/SciPostPhys.7.2.024,Coppola_2022,PhysRevLett.120.133601}, one considers continuous  monitoring of the local density of fermions, $n_i=c^\dagger_i c_i$. 
In this context, the measurement induces decoherence and the state averaged over all possible measurement outcomes, $\rho_t$, is generally mixed. Its evolution is governed by the Lindblad master equation \cite{10.21468/SciPostPhys.7.2.024}
\begin{equation}
\dot{\rho}_t = -i \left[H,\rho_t \right] + \frac{\gamma}{2} \sum_{i=1}^L \left[\left[n_i,\rho_t\right],n_i\right]\, .
        \label{Lindblad}
\end{equation}
Single realizations of the monitored dynamics are instead modeled through appropriate stochastic Schr\"odinger equations. The latter describe the evolution of the pure system state $\ket{\psi_t}$ conditional on the outcome of the continuous measurement. In contrast to the average evolution in Eq.~\eqref{Lindblad}, these dynamics are strongly dependent on the details of the monitoring and there can be different measurement processes giving rise to master equation \eqref{Lindblad}. 
One such process  is associated with the stochastic Schr\"odinger equation \cite{10.21468/SciPostPhys.7.2.024}
\begin{equation}
      {\rm d}|\psi_t\rangle\!=\! -iH {\rm d}t|\psi_t\rangle   + \sum_{i=1}^{L} \!\left(\sqrt{\gamma_{\rm I}} \Delta_t n_i {\rm d} W_t^{i}-\frac{\gamma_{\rm I}}{2}\Delta_t^2 n_i  {\rm d}t\right)\!|\psi_t\rangle\, ,
    \label{Unravelling1}
\end{equation}
which provides the increment of the system state, ${\rm d}\ket{\psi_t}$, during an infinitesimal time interval ${\rm  d}t$. The term ${\rm d}W_t^i$ is the increment of a  standard Wiener process, obeying  $\mathbb{E}[{\rm d}W_t^i {\rm d}W_t^j]=\delta_{ij}{\rm d}t$ and $\mathbb{E}[{\rm d}W_t^i]=0$, with $\mathbb{E}$ denoting expectation over the  processes. We have further defined the operator $ \Delta_t n_i = \left[n_i-\langle n_i \rangle_t\right]$. The process in Eq.~\eqref{Unravelling1} is used to model, for instance, homodyne-detection measurements \cite{PhysRevLett.120.133601}, with $\gamma_{\rm I}$ encoding the monitoring rate. 
For a large ratio $\gamma_{\rm I}/\lambda$, the average entanglement entropy $\bar{S}_\infty$ of the process saturates to a value which does not depend on system size, indicative of an area-law behavior as shown in Fig.~\ref{Fig2}(a) \cite{Alberton_2021,10.21468/SciPostPhys.7.2.024}. Conversely, for small values $\gamma_{\rm I}/\lambda$, the system enters a phase with subextensive stationary entanglement entropy, growing as the logarithm of the system size \cite{Alberton_2021,Coppola_2022}. The unraveling in Eq.~\eqref{Unravelling1}, therefore, does not feature volume-law entanglement \cite{Alberton_2021,10.21468/SciPostPhys.7.2.024}.

However, one can construct a stochastic process, associated with Eq.~\eqref{Lindblad}, for which volume-law entanglement is possible. This is the case when considering, for instance, the stochastic equation \cite{10.21468/SciPostPhys.7.2.024} 
\begin{equation}
     {\rm d}\ket{\psi_t} = -iH{\rm d}t\ket{\psi_t}   - \sum_{i=1}^{L} \left(\frac{\gamma_{\rm II}}{2}n_i {\rm d}t +i \sqrt{\gamma_{\rm II}}n_i {\rm d}V_t^{i} \right)\ket{\psi_t}\, ,
    \label{Unravelling2}
\end{equation}
with ${\rm d}V_t^{i}$ being a standard Wiener increment. 
Such a process is also related to homodyne detection \cite{PhysRevLett.120.133601} and $\gamma_{\rm II}$ is the monitoring rate. Due to the imaginary unit in front of the noise terms, the process in Eq.~\eqref{Unravelling2} is  generated by a stochastic Hamiltonian and has effectively no nontrivial non-Hermitean component. 
Essentially, Eq.~\eqref{Unravelling2} describes the system dynamics in the presence of a fluctuating on-site potential  described by independent Wiener increments.   
As clearly displayed in Fig.~\ref{Fig2}(b), in this case bipartite entanglement 
grows as $\bar{S}_t\sim \sqrt{\lambda t}$ \cite{young2024,10.21468/SciPostPhys.7.2.024} and approaches a saturation value $\bar{S}_\infty$ displaying, on average, the same volume-law behavior shown by the deterministic unitary dynamics $\gamma = 0$ (see Fig. \ref{SuppFig2} in Appendix \hyperref[sec:appendixB]{B}  for a comparison).

\section{Interpolating between unravelings}
Evidently, the stochastic processes in Eq.~\eqref{Unravelling1} and in Eq.~\eqref{Unravelling2} result in qualitatively different scaling of entanglement.
In the following, we want to explore whether interpolating between the two, while keeping the measurement rate fixed, can result in an entanglement phase transition. By examining  Fig.~\ref{Fig2}(a) and Fig.~\ref{Fig2}(b), one may  expect a transition from area-law to volume-law behavior. However, as we will show, at most logarithmic entanglement scaling is possible. 

Starting point is the stochastic Schr\"odinger equation \begin{equation}
\begin{split}
    {\rm d} \ket{\psi_t} &\!=\! \biggl[-iH - \sum_{i=1}^L\left(\frac{1-\alpha}{2}\gamma\Delta_t^2  n_i+\frac{\alpha \gamma}{2}n_i  \right) {\rm d}t \\
    & \!+\sum_{i=1}^L\left( \sqrt{(1-\alpha)\gamma}\Delta_t  n_i {\rm d}W_t^{i} -i\sqrt{\alpha \gamma}n_i {\rm d}V_t^{i} \right) \biggr]\!\ket{\psi_t}  \, ,
    \end{split}\label{Unravelling3}
\end{equation}
which mixes the dynamical contributions from Eq.~\eqref{Unravelling1} and Eq.~\eqref{Unravelling2}, by setting $\gamma_{\mathrm{I}} =(1-\alpha)\gamma$ and $\gamma_{\mathrm{II}} = \alpha \gamma$. We consider  $\lambda=\gamma/2$ throughout our investigation. The parameter $\alpha \in \left[0,1\right]$ plays the role of an unraveling parameter which allows one to generate a family of unravelings, interpolating between the two processes introduced above. Note that, upon averaging over all possible realizations of the stochastic increments ${\rm d}W_t^i$ and ${\rm d}V_t^i$,  Eq.~\eqref{Unravelling3} reproduces the Lindblad evolution in Eq.~\eqref{Lindblad} for any allowed value of $\alpha$.

Before presenting our results for the different stochastic processes generated by Eq.~\eqref{Unravelling3}, we briefly discuss the numerical procedure we exploit  \cite{10.21468/SciPostPhys.7.2.024,Alberton_2021,piccitto2022} (for a detailed discussion see Appendix \hyperref[sec:appendixA]{A}). Since the  dynamics in Eq.~\eqref{Unravelling3} is implemented by a quadratic operator which conserves  the total number of particles, we can write the state of the system, at any time $t$, as \begin{equation}
    |\psi_t \rangle = \prod_{k=1}^N \left( \sum_{j=1}^L U_{jk}(t) c^{\dagger}_j\right) |0 \rangle.
    \label{efficientstate}
\end{equation}
Here, $N$ is the number of fermionic particles (in our case $N=L/2$) and $U$ is an $L\times N$ isometry, $U^\dagger U=\mathds{1}_{ N}$ where $\mathds{1}_{ N}$ is the identity matrix of dimension $N$. To analyze the dynamics of the quantum state, it is  sufficient to understand how the matrix $U(t)$ evolves. By considering  Eq.~\eqref{Unravelling3} and by discretizing time, we can find the operator $K_t$ such that $\ket{\psi_{t+{\rm d}t}}\approx e^{K_t}\ket{\psi_t}$. One can check that $K_t\ket{0}=0$ and that $K_t$ is quadratic. As such, the evolution of $U(t)$ can be obtained by calculating the linear combinations of creation operators generated by $e^{K_t }c_j^\dagger e^{-K_t}=\sum_{m=1}^L R_{jm} c_m^\dagger$. This provides an update rule for the matrix $U(t)\to U(t+{\rm d}t)$ and we set $\gamma {\rm d}t=0.1$ in our numerical simulations  (see also Fig. \ref{SuppFig1} in Appendix \hyperref[sec:appendixB]{B}  for an analysis with $\gamma {\rm d}t=0.05$). The isometry  $U(t)$ gives access to the fermionic single-particle correlation function $D_{mn}(t) = \langle c^{\dagger}_m c_n \rangle_t = \left[U(t) U^{\dagger}(t)\right]^*_{mn}$, from which we can calculate the entanglement entropy \cite{Peschel_2003,Peschel_2009,10.21468/SciPostPhys.7.2.024,Alberton_2021} (see details in Appendix \hyperref[sec:appendixA]{A}). We average this quantity over several realizations of the stochastic processes and estimate its stationary behavior by integrating over a finite window at large times. By construction, the entanglement entropy behaves as shown in Fig.~\ref{Fig2}(a,b), for the limiting cases $\alpha = 0$ and $\alpha = 1$, showing area-law and volume-law entanglement, respectively. 

\begin{figure}[t]      
        \includegraphics[width = \columnwidth]{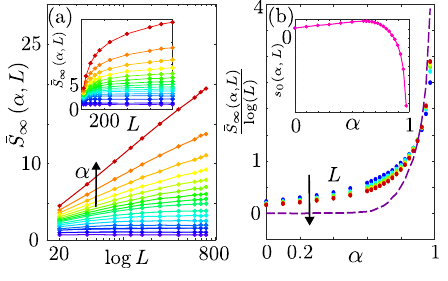}        \caption{\textbf{Unraveling-induced entanglement phase transition. } (a) Stationary values of the entanglement entropy $\bar{S}_\infty(\alpha,L)$ as a function of the logarithm of the system size, for different values of $\alpha$. The inset shows the data in linear scale. For small parameter $\alpha$, $\bar{S}_\infty(\alpha,L)$ 
        exhibits an area-law behavior. Upon increasing $\alpha$, $\bar{S}_\infty(\alpha,L)$ starts to display a subextensive logarithmic growth with the system size. (b) Estimate of the order parameter $c(\alpha)$ [see Eq.~\eqref{Entropy_form}], obtained as $\bar{S}_\infty(\alpha,L)/\log L$, for $L \in \left\lbrace 120,250,400,600,700,800\right\rbrace$. The dashed line represents  $c(\alpha)$ obtained from a fit of Eq.~\eqref{Entropy_form} to the data of the same system sizes (or the six largest ones). The inset shows instead the fit of the residual entropy $s_0(\alpha)$. We have considered $\lambda=\gamma/2 $ and the average is performed over $100$ realizations of the process. }
        \label{Fig3}
\end{figure}

\section{Unraveling phase transition}
In the regime $0<\alpha<1$, we see that no volume-law entanglement scaling is possible [cf.~Fig.~\ref{Fig3}(a)]. Nonetheless, two different  
regimes  emerge. For $\alpha$ smaller than a critical value $\alpha_{\rm c}$, the stochastic process features an area-law entanglement behavior, $\bar{S}_\infty \propto O(1)$ as in the case of $\alpha=0$.  On the other hand, for $\alpha >\alpha_{\rm c}$ after a transient regime displaying a logarithmic entanglement growth in time  $\propto \log\left(\gamma t\right)$  [see Fig.\ref{SuppFig2}(a) in Appendix \hyperref[sec:appendixB]{B}], the system approaches a stationary behavior proportional to the logarithm of the subsystem size,  $\bar{S}_\infty \propto \log L $. The latter is reminiscent of the one observed in Ref.~\cite{Alberton_2021} upon variations of the measurement rate.  

To better understand the transition between the two phases and to get insights on the value of $\alpha_{ \rm c}$, we assume the following scaling form for the stationary entanglement entropy, as a function of $\alpha$ and $L$, \cite{Alberton_2021}
\begin{equation}
    \bar{S}_\infty\left(\alpha,L\right) \approx  c(\alpha)\log L+s_0\left(\alpha\right) \, .
    \label{Entropy_form}
\end{equation}
Within the above expression, the coefficient $c(\alpha)$ acts as an order parameter for the entanglement transition. Indeed, when the system is found in the area-law phase, one expects to observe $c(\alpha)\approx 0$, while the logarithmic phase should be characterized by a finite value $c(\alpha)> 0$. The term $s_0(\alpha)$ represents a constant offset to the entanglement entropy. 
The behavior of the order parameter, as  approximated by $c(\alpha)\approx \bar{S}_\infty\left(\alpha,L\right)/\log L$ for increasing values of $L$, is shown in Fig.~\ref{Fig3}(b). We observe a region in which the ratio $\bar{S}_\infty\left(\alpha,L\right)/\log L$ tends to zero, thus witnessing area-law phase. However, for values of $\alpha$ larger than a critical one, $\alpha_{\rm c}$, the ratio increases with $L$ and appears to approach a finite size-independent value. In Fig.~\ref{Fig3}(b), we also show, for comparison, the value of $c(\alpha)$ as obtained from a fit of Eq.~\eqref{Entropy_form}. The inset provides instead a fit for the  area-law term $s_0(\alpha)$. 

\begin{figure}[t]
\includegraphics[width = \columnwidth]{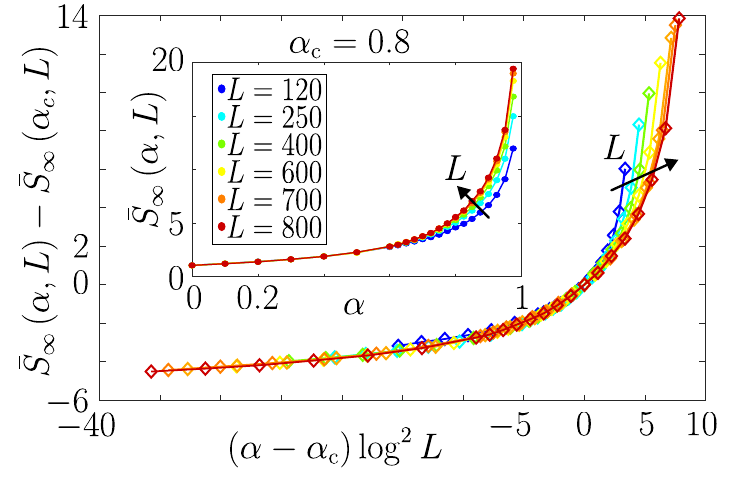}    \caption{\textbf{Finite-size scaling. } Finite-size scaling of the stationary entanglement entropy, assuming BKT universality (see main text and Ref.~\cite{Alberton_2021}). The difference $\bar{S}_\infty(\alpha,L)-\bar{S}_\infty(\alpha_{\rm c},L)$ is plotted against the quantity $(\alpha-\alpha_{\rm c})\log^2L$, for $\alpha_{\rm c}=0.8$. The inset shows the bare data for $\bar{S}_\infty(\alpha,L)$. As in the other plots, we have considered $ \lambda=\gamma/2$ and the average is performed over $100$ realizations of the process. }
    \label{Fig4}
\end{figure}
 
The presence of an extended logarithmic regime, reminiscent of a conformally invariant critical phase in purely Hamiltonian dynamics \cite{Holzhey_1994,PhysRevLett.90.227902,Calabrese_2009}, is suggestive of Berenzinskii-Kosterlitz-Thouless (BKT) universal behavior \cite{PhysRevD.80.125005} close to the transition point \cite{Alberton_2021}. Under this assumption, we can estimate the critical unraveling parameter which separates the two entanglement regimes. First of all, the value of $\alpha$ at which the $s_0(\alpha)$ changes sign, from positive to negative, may provide an estimate for the critical unraveling parameter $\alpha_{\rm c}$ \cite{Alberton_2021}. To consolidate this estimate, which from the inset of Fig.~\ref{Fig3}(b) appears to be roughly $\alpha_{\rm c}\approx 0.8$, we explore a finite-size scaling analysis for $\bar{S}_\infty(\alpha,L)$ \cite{Alberton_2021,PhysRevA.86.043629,PhysRevB.55.R11949}. We consider the function $\bar{S}_{\infty}\left(\alpha, {L}\right) - \bar{S}_{\infty}\left(\alpha_c, {L}\right)$ and plot it as a function of the quantity $\left(\alpha - \alpha_c\right) \log^2 L$ for different values of $\alpha_{\rm c}$. When considering the exact value of the critical parameter, one would expect to observe a collapse of all data points onto a unique scaling function. In Fig.~\ref{Fig4} we report the scaling that we have obtained by assuming $\alpha_{\rm c}=0.8$. For the latter value, the finite-size scaling works reasonably well, given also unavoidable finite-size effects, suggesting that the transition is within the BKT universality.  We also report a more systematic analysis of the scaling in Appendix \hyperref[sec:appendixB]{B} [cf.~Fig.~\ref{SuppFig3}]  leading to a value $\alpha_{\rm c}=0.875$. The latter, given finite-size effects and our finite data set, is in reasonable agreement with the value providing the scaling in Fig.~\ref{Fig4}.
The inset of Fig.~\ref{Fig4} shows the bare data for $\bar{S}_\infty\left(\alpha,L\right)$ as a function of $\alpha$.

\section{Conclusions} We have considered a family of continuous-monitoring processes identified by an unraveling parameter $\alpha$. We have shown that varying the measurement process (achieved by varying $\alpha$) can have a dramatic impact on the spreading of entanglement in the considered noninteracting tight-binding model. When the unraveling parameter reaches a critical value, the system transitions from a phase characterized by an area-law to one exhibiting logarithmic entanglement behavior. The latter phase is reminiscent of emergent conformal invariance in unitary systems and the  entanglement phase transition indeed appears to belong to the BKT universality \cite{Alberton_2021}. 
It would be interesting to study the entanglement phase diagram in the $\gamma-\alpha$ plane \cite{PhysRevLett.128.243601}. We expect that for sufficiently small $\gamma$ the system will always be found in the logarithmic phase. On the other hand, we expect that for large $\gamma$, the value of $\alpha_{\rm c}$, determining the onset of the logarithmic phase, moves towards $\alpha_{\rm c}\to1$. 
 We note that the critical value $\alpha_{\rm c}$ that we estimate for our parameters seems to be in line with what could be obtained by considering only the (adimensional) effective measurement rate $(\gamma/\Omega)(1-\alpha)$ and comparing it with the critical value of Ref.~\cite{Alberton_2021}. To understand whether this is just a coincidence of whether the rate $\gamma_{\rm II}$ is really ``irrelevant" for the transition, one would need indeed to derive the full phase diagram previously mentioned.

The nature and even the existence of a phase with logarithmic entanglement scaling is currently debated \cite{Alberton_2021,minato2022,Coppola_2022,PhysRevX.11.041004,Poboiko_2023,Poboiko_2024,Fava_2023,starchl2024}. Arguments based on symmetry considerations suggest that such logarithmic phase merely appears as a transient, eventually saturating at a finite area-law value for the entanglement in the thermodynamic limit $L\to\infty$ \cite{Poboiko_2023}. If this is indeed the case, our numerical observations would reveal a sharp crossover in the entanglement behavior, as a function of the unraveling, for the analyzed system sizes.

In Ref.~\cite{PhysRevLett.128.243601} an unraveling transition was observed in a fully random quantum circuit, where one of the two ``extremal" unravelings could be absorbed in the random Hamiltonian ensemble. In contrast, our system features a deterministic Hamiltonian evolution and two unravelings both substantially altering  the unitary dynamics. As a result, the entanglement phase transition we observe cannot be directly mapped onto a ``standard" measurement-induced one \cite{PhysRevLett.128.243601}.  It would be interesting to explore whether similar phenomenology might arise in random circuits for quantum communication (see, e.g., Ref.~\cite{kelly2023}), for example considering different unravelings of the same quantum channel.

Signatures of our unraveling-induced entanglement phase transition may be observed in experiments with cold-atoms \cite{PhysRevLett.120.133601}, for instance by investigating the behavior of many-body correlations \cite{Alberton_2021}. Experimentally, different unravelings may be obtained by varying the quadrature component of the monitored output field \cite{PhysRevLett.120.133601}. This possibility may allow for controlling entanglement in many-body states via continuous monitoring, which is potentially relevant for tailoring quantum correlations in noisy devices.

\acknowledgments
We thank Marcel Cech, Gabriele Perfetto and Chris Nill for fruitful discussions. We are further grateful to Xhek Turkeshi for useful discussions and for drawing our attention to recent literature.
We acknowledge funding from the Deutsche Forschungsgemeinschaft (DFG, German Research Foundation) under Project No.~435696605 and through the Research Unit FOR 5413/1, Grant No.~465199066, through the Research Unit FOR 5522/1, Grant No.~499180199 and from the state of Baden-W\"urttemberg through bwHPC grant no. INST 40/575-1 FUGG (JUSTUS 2 cluster). This project has also received funding from the European Union’s Horizon Europe research and innovation program under Grant Agreement No.~101046968 (BRISQ). F.C.~is indebted to the Baden-W\"urttemberg Stiftung for the financial support of this research project by the Eliteprogramme for Postdocs. This work was supported by the QuantERA II programme (project CoQuaDis, DFG Grant No. 532763411) that has received funding from the EU H2020 research and innovation programme under GA No. 101017733.





\newpage
\appendix
\section{Numerical implementation} \label{sec:appendixA}
\phantomsection
\setcounter{equation}{0}

\setcounter{figure}{0}
\setcounter{table}{0}
\makeatletter
\renewcommand{\theequation}{A\arabic{equation}}
\renewcommand{\thefigure}{A\arabic{figure}}
\renewcommand{\thetable}{A\arabic{table}}

\makeatletter
\renewcommand{\theequation}{A\arabic{equation}}
\renewcommand{\thefigure}{A\arabic{figure}}
\renewcommand{\thetable}{A\arabic{table}}
In this section, we provide details on the numerical method we exploit to simulate the diffusive quantum trajectories discussed in the main text. 

\subsection{Representation of the state and covariance matrix}
The first step is to understand how the representation of the state given in Eq.~\eqref{efficientstate} can be used to derive the single-particle covariance matrix \(D_{mn} = \langle c_m^\dagger c_n \rangle\). 

We start by showing that a state given in the form 
$$
\ket{\psi}=\prod_{k=1}^N \left( \sum_{j=1}^L U_{jk} c^{\dagger}_j\right) |0 \rangle\, ,
$$
with $U$ being an isometry such that $U^\dagger U=\mathds{1}_{N}$, is normalized. The idea is the following. Since $U$ is an isometry, we can write it as 
$$
U=\begin{pmatrix}
\vec{v}_1 & \vec{v}_2  & \cdots & \vec{v}_N  \\
\end{pmatrix} \, ,
$$
with the $L$-dimensional vectors $\vec{v}_i$ such that $ \vec{v}_i^*\cdot\vec{v}_j= \delta_{ij}$. This means that these $N$ vectors form an incomplete basis of $\mathbb{C}^L$. Completing the basis introducing arbitrary orthogonal vectors $\left\lbrace\vec{v}_{N+1}\dots \vec{v}_L\right\rbrace$, which are now such that $ \left( \vec{v}_i^*,\vec{v}_j \right) = \delta_{ij}$, for all $i,j=1,2,\dots L$, we can promote the isometry to a unitary operator 
$$
\tilde{U}=\begin{pmatrix}
\vec{v}_1 & \vec{v}_2  & \cdots & \vec{v}_N \dots \vec{v}_L \\
\end{pmatrix} \, ,
$$
and we can also write the state as
$$
\ket{\psi}=\prod_{k=1}^N \left( \sum_{j=1}^L \tilde{U}_{jk} c^{\dagger}_j\right) |0 \rangle\, .
$$
Note indeed that the arbitrarily chosen vectors do not appear in the above summation. 

Through the unitary matrix $\tilde{U}$, we can define new fermionic modes as 
\begin{align}
C^{\dagger}_k = \sum_{j=1}^L \tilde{U}_{jk} c^{\dagger}_j, \, \qquad 
C_k = \sum_{j=1}^L \tilde{U}^*_{jk}c_j\, ,
\label{Trafo}
\end{align}
which obey standard canonical anti-commutation relations, $\left\{C_k,C_h^\dagger \right\}=\delta_{kh}$. This also allows us to write the state as $\ket{\psi}=C_N^\dagger C^{\dagger}_{N-1}\dots C_2^\dagger C_1^\dagger\ket{0}$, which is thus clearly normalized. Considering also the inverse transformations 
\begin{align*}
c^{\dagger}_m = \sum_{k=1}^L \tilde{U}^*_{mk} C^{\dagger}_k, \,\quad 
c_m = \sum_{k=1}^L \tilde{U}_{mk} C_k, 
\end{align*}
we can calculate the matrix $D$ as 
\begin{align*}
D_{mn} &=\langle \psi_t | c^{\dagger}_m c_n | \psi_t \rangle \\ &= \sum_{j,k=1}^L \langle 0| C_1 \dots C_N \tilde{U}^*_{mj} \tilde{U}_{nk} C^{\dagger}_j C_k C^{\dagger}_N \dots  C^{\dagger}_1 |0 \rangle. 
\end{align*}
With regard to the sum in the above equation, we can immediately see that, if $k>N$ or if $j>N$ the contribution to the sum is zero. Moreover, even for $k,j\le N$ we only have a nonzero contribution when $j=k$. As such, we find 
\begin{align*}
D_{mn}=
\sum_{k=1}^N \tilde{U}^*_{mk}\tilde{U}_{nk}=\sum_{k=1}^N {U}^*_{mk}{U}_{nk}=\left[{U} {U}^{\dagger}\right]^*_{mn}.
\end{align*}
In the above equation, we have exploited the fact that the vectors added to the isometry in order to obtain a proper unitary matrix are irrelevant in the summation. 
By calculating this quantity for the entire system, we are able obtain the eigenvalues of the correlation function of subsystem $A$ by reading off the submatrix of $D$ for $i,j \in A$.

\subsection{Time evolution of the state}
Given that the generator of the dynamics is quadratic and  conserves the number of particles in the system, the parametrization in Eq.~\eqref{efficientstate} of the main text is valid at any time $t>0$. Hence, to investigate dynamical  properties of the system,  it is sufficient to understand how the matrix $U(t)$ evolves in time. 

To this end, we first calculate the operator $K_t$ such that the equation of the increment in Eq.~\eqref{Unravelling3} is equivalent to the update rule 
$$
\ket{\psi_{t+{\rm d}t}}\approx e^{K_t}\ket{\psi_t}. 
$$
By exploiting Ito's Lemma we find that this is achieved by considering 
\begin{align}
K_t=&-iH{\rm d}t-\sum_{i=1}^L\langle n_i \rangle_t \left( \sqrt{\gamma_{\mathrm{I}}}{\rm d}W_t^{i}+\gamma_{\mathrm{I}} \langle n_i \rangle_t {\rm d}t \right) \nonumber\\&+\sum_{i=1}^Ln_i\left(\gamma_{\mathrm{I}}\left(2\langle n_i\rangle_t-1\right){\rm d}t-i\sqrt{\gamma_{\mathrm{II}}}{\rm d}V_t^{i}+\sqrt{\gamma_{\mathrm{I}}}{\rm d}W_t^{i}\right)   \, .\label{Timeevolutor3}
\end{align}
In the numerical implementation of the above equation, we consider a small, but finite, ${\rm d}t$ and we generate Wiener processes by considering ${\rm d}W_t^{i}$ and ${\rm d}V_t^{i}$ to be random Gaussian numbers with average zero and variance $\sqrt{{\rm d}t}$. For our numerical simulations, we use $\gamma{\rm d}t=0.1$.It is notable, that while the trotterization \eqref{Timeevolutor3} might hold for smaller $\rm{d}t$ it might not work for the one chosen. In order reconsidering this we resimulated the data setting $\rm{d}t= 0.05$ and recovered the same expected behaviour  for the $\rm{d}t$ chosen as seen in Fig. \ref{SuppFig3}.

Additionally, we note that $K_t\ket{0}=0$ so that we can write 
\begin{align*}
    e^{K_t}\ket{\psi_t}&=e^{K_t}\prod_{k=1}^N \left( \sum_{j=1}^L U_{jk}(t) c^{\dagger}_j\right) |0 \rangle\\&=\prod_{k=1}^N \left( \sum_{j=1}^L U_{jk}(t) e^{K_t}c^{\dagger}_j e^{-K_t}\right) |0 \rangle\, .
\end{align*}
By direct calculation, we find that 
$$
e^{K_t}c^{\dagger}_j e^{-K_t}\approx \sum_{k=1}^L\left(M e^{-ihdt} U(t)\right)_{kj}c_k^{\dagger}\, ,
$$
which defines the update rule for $U(t)$ as 
\begin{align*}
&U_{jk}(t+{\rm d}t)=\left[M e^{-i h{\rm d}t}U(t)\right]_{jk}\,, \\ & \mbox{with}\quad M_{jk} = \delta_{jk} e^{\left[\gamma_{\mathrm{I}}\left(2\langle n_k\rangle_t-1\right){\rm d}t-i\sqrt{\gamma_{\mathrm{II}}}{\rm d}V_t^{k}+\sqrt{\gamma_{\mathrm{I}}}{\rm d}W_t^{k}\right]}\, .    
\end{align*}
To preserve the isometry property of $U(t+\rm{d}t)$, despite the discretized evolution being non-unitary, we perform a QR decomposition of $U(t+\rm{d}t)$ and redefine $U(t+\rm{d}t) = Q$ \cite{Alberton_2021,10.21468/SciPostPhys.7.2.024}. The isometry for the initial state is given by 
$$
U(0)=U_{jk}(0) = \delta_{j,2k}.
$$

To conclude, we mention that the entanglement entropy of a subsystem $A$ consisting of the first $\ell$ fermionic sites can be computed by taking the eigenvalues $\lambda_j^{(A)}$ of the covariance matrix $D_A$ for subsystem $A$ and calculating \cite{Peschel_2003,Peschel_2009}
\begin{align}
    S= - \sum_{j=1}^\ell \left[\lambda_j^{(A)}\log\left(\lambda_j^{(A)}\right) + \left(1-\lambda_j^{(A)}\right)\log\left(1-\lambda_j^{(A)}\right)\right].
    \label{Entropy}
\end{align}

\section{Additional results} \label{sec:appendixB}
\renewcommand{\theequation}{B\arabic{equation}}
\renewcommand{\thefigure}{B\arabic{figure}}
\renewcommand{\thetable}{B\arabic{table}}

\makeatletter
\renewcommand{\theequation}{B\arabic{equation}}
\renewcommand{\thefigure}{B\arabic{figure}}
\renewcommand{\thetable}{B\arabic{table}}
In this section, we provide additional results, including an explicit fit of the Ansatz to the generated data, additional data for a smaller time-increment, plots regarding both volume-law saturation values and the initial entropy dynamics as well as a more systematic approach to the finite-size scaling.

\subsection{Results for a smaller time-step size}

\begin{figure}[H]   
  \centering
  \includegraphics[width=\columnwidth]{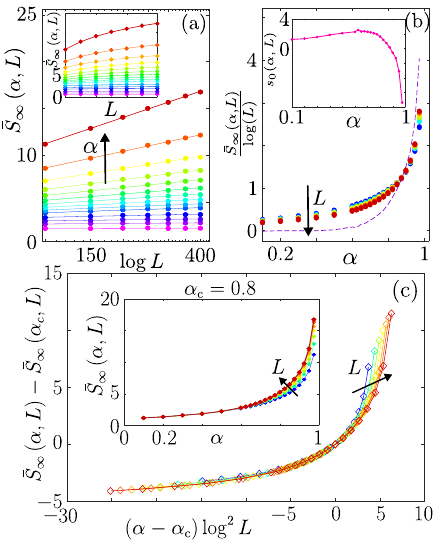}
  \caption{\textbf{Unraveling-induced entanglement phase transition for $\gamma \rm{d}t = 0.05.$ } (a) Stationary values of the entanglement entropy $\bar{S}_\infty(\alpha,L)$ as a function of the logarithm of the system size, for different values of $\alpha$. The inset shows the data in linear scale. For small parameter $\alpha$, $\bar{S}_\infty(\alpha,L)$ 
        exhibits an area-law behavior. Upon increasing $\alpha$, $\bar{S}_\infty(\alpha,L)$ starts to display a subextensive logarithmic growth with the system size. (b) Estimate of the order parameter $c(\alpha)$ [see Eq.~\eqref{Entropy_form}], obtained as $\bar{S}_\infty(\alpha,L)/\log L$, for $L \in \left\lbrace 100,150,200,250,300,350,400\right\rbrace$. The dashed line represents  $c(\alpha)$ from a fit of Eq.~\eqref{Entropy_form} using the largest four system sizes considered in panel (a). The inset shows instead the fit of the residual entropy $s_0(\alpha)$. We have considered $\lambda=\gamma/2 $ and the average is performed over $100$ realizations of the process.
        (c) Finite-size scaling of the stationary entanglement entropy, assuming BKT universality (see main text and Ref.~\cite{Alberton_2021}). The difference $\bar{S}_\infty(\alpha,L)-\bar{S}_\infty(\alpha_{\rm c},L)$ is plotted against the quantity $(\alpha-\alpha_{\rm c})\log^2L$, for $\alpha_{\rm c}=0.8$. The inset shows the bare data for $\bar{S}_\infty(\alpha,L)$. We have considered $\lambda=\gamma/2 $ and the average is performed over $100$ realizations of the process.}
        \phantomsection
  \label{SuppFig1}
\end{figure}

In this section we show results for $\gamma\rm{dt}=0.05$ in order to rule out the possibility that the behavior observed in the main text is influenced by a too  large Trotterization step. These results are shown in Fig.~\ref{SuppFig1} and display an entanglement entropy behavior compatible with the one discussed in the main text for $\gamma \rm{d}t = 0.1$. 

\subsection{Volume law, initial entropy dynamics and collapse of stationary entropies}
\begin{figure}[t]   
\centering
\includegraphics[width=\columnwidth]{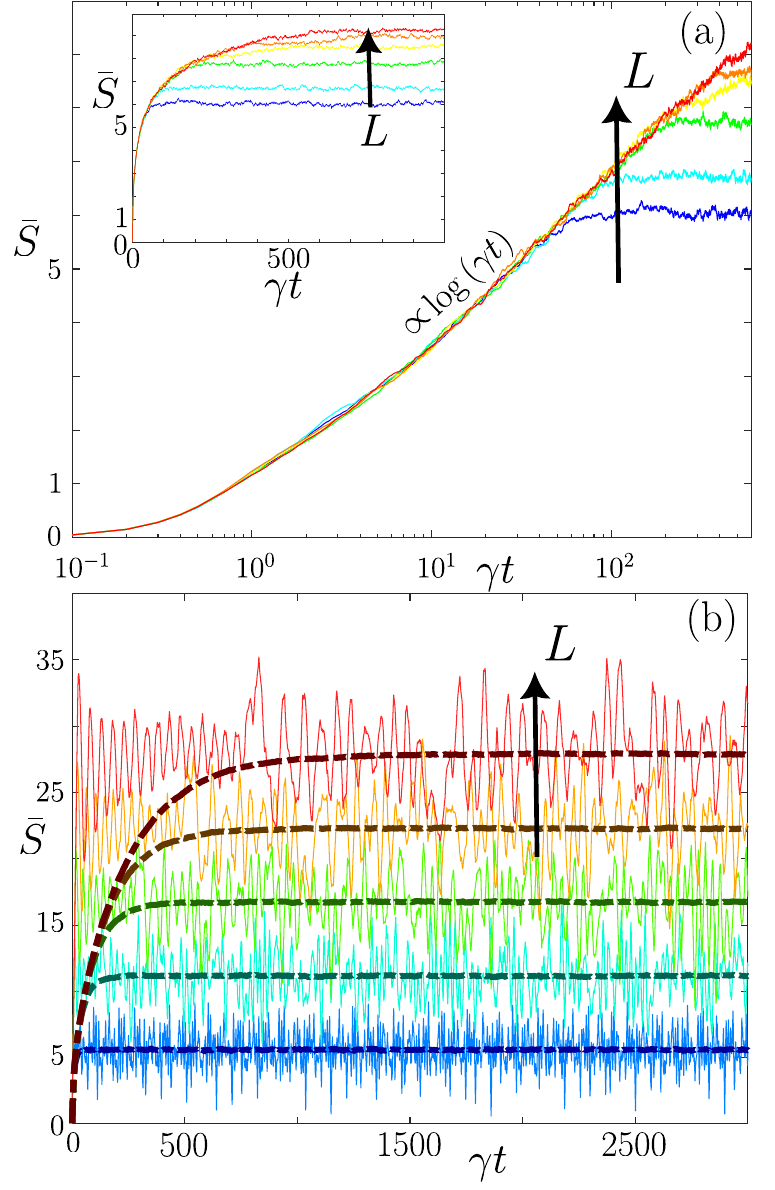}
  \caption{\textbf{Entanglement entropy dynamics.} (a) Entanglement entropy averaged over 100 realizations of the process in the logarithmic phase ($\alpha=0.9$) with a logarithmic x-axis. The plot shows the initial logarithmic-in-time growth of entanglement. The inset displays  the bare  data for $\alpha = 0.9$. (b) Entanglement entropy, $\bar{S}$, averaged over $100$ realizations of the stochastic  process in Eq.~\eqref{Unravelling3} for $\alpha = 1$. Dotted lines represent the entanglement entropy for $\alpha = 1$ while the connected lines represent the unitary case $\gamma = 0$. }
  \phantomsection
  \label{SuppFig2}
\end{figure}

In order to get additional insight onto the entropy dynamics in our model, we also briefly   study the transient behavior in the logarithmic phase. While it is known that  in the volume law the entanglement entropy grows as $\bar{S}_t \propto \sqrt{t}$ \cite{young2024,10.21468/SciPostPhys.7.2.024}, we find that in the logarithmic regime it grows as $\propto \log\left(t\right)$, as shown in Fig.~\ref{SuppFig2}(a), before saturating to  a value $\propto \log\left(L\right)$.\\
We also look more in detail at the stationary entanglement behavior for two volume-law cases considered: the case of the unravelling with $\alpha = 1$ and that of the purely unitary dynamics, or $\gamma = 0$ case. We observe that both scenarios  yield not only the same volume law character, but the same stationary entanglement as displayed in Fig.~\ref{SuppFig2}(b). However, the $\gamma = 0$ scenario exhibits an oscillatory behavior around the stationary value while the $\alpha = 1$ scenario approaches it directly and remains stable.

\begin{figure}[H]  
\centering
  \includegraphics[width=\columnwidth]{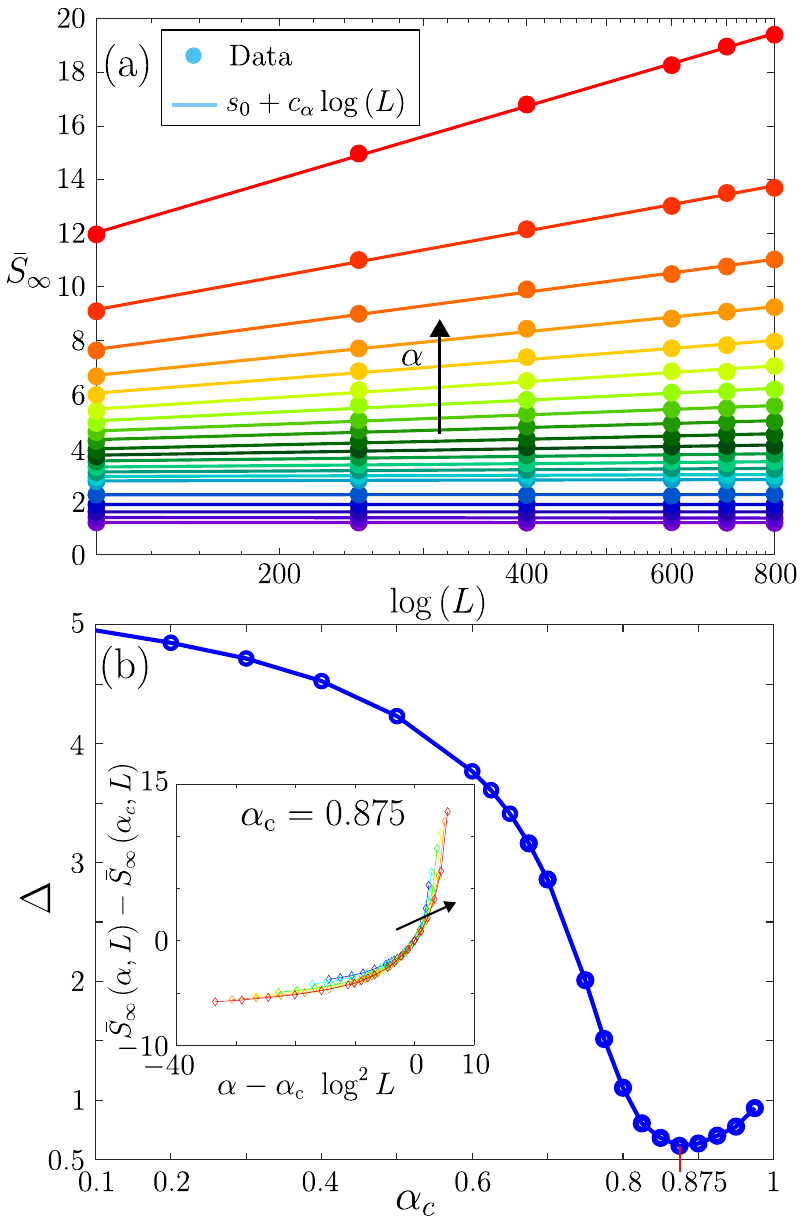}
  \caption{\textbf{Fit of the Ansatz and systematic finite size scaling.} (a) Data of the stationary entropy fitted with \eqref{Entropy_form} for all values of $\alpha$ over the logarithm of the system size $L$. The dots represent the numerically obtained data points, and the lines the fit. 
  (b) Total distance $\Delta(y_i,y_j)$ of the curves for different critical interpolation values $\alpha_{\rm c}$ extracted assuming BKT universality as exemplary shown in \ref{Fig4}. where $\left\lbrace y_i\right\rbrace$ are the interpolated y-values $\bar{S}_\infty(\alpha,L)-\bar{S}_\infty(\alpha_{\rm c},L)$ for each critical parameter $\alpha_c$ onto a general grid of x values. The Inset shows the finite size scaling of the stationary entanglement entropy, assuming BKT universality.}
  \phantomsection
  \label{SuppFig3}
\end{figure}
Finally, we also provide here a more ``objective" approach to the estimate of the critical parameter $\alpha_{\rm c}$, inspired by what has been done in Ref.~\cite{PhysRevB.101.060301}. To this end, we introduce a function which considers  the distances between all pairs of curves obtained by rescaling the numerical data for the different system sizes. Such  function solely depends on the free parameter $\alpha_{\rm c}$.  More precisely, we consider the normalized distance between curves
\begin{align}
    \Delta = \sum_{i,j ,i\neq j} \frac{|y_i-y_j|}{|y_i|+|y_j|}.
    \label{Distance}
\end{align}
Here, the set $\left\lbrace y_i \right\rbrace $ represents the $\ y $-values, defined as $ \bar{S}_\infty(\alpha, L) - \bar{S}_\infty(\alpha_{\rm c}, L) $ as in Fig.~\ref{Fig4}, obtained by interpolation onto a general grid of $ x $-values, defined as $\left(\alpha-\alpha_{\mathrm{c}}\right)\log^2L$ in order to compare curves obtained for different $L$. \\

After defining the function, we want to find its minimum upon varying the free parameter $\alpha_{\rm c}$. Such a minimum indicates the value for which the scaling is optimal. According to Fig.~\ref{SuppFig3}(b), the best collapse is obtained in the region around $\alpha_{\rm c} \approx 0.875$.  Such a value is in line, given the finite-size effects that we have and the interpolation we have to perform due to our finite data set, with the one reported in the main text. 
\newpage
\bibliography{biblio}

\begin{thebibliography}{62}%
\makeatletter
\providecommand \@ifxundefined [1]{%
 \@ifx{#1\undefined}
}%
\providecommand \@ifnum [1]{%
 \ifnum #1\expandafter \@firstoftwo
 \else \expandafter \@secondoftwo
 \fi
}%
\providecommand \@ifx [1]{%
 \ifx #1\expandafter \@firstoftwo
 \else \expandafter \@secondoftwo
 \fi
}%
\providecommand \natexlab [1]{#1}%
\providecommand \enquote  [1]{``#1''}%
\providecommand \bibnamefont  [1]{#1}%
\providecommand \bibfnamefont [1]{#1}%
\providecommand \citenamefont [1]{#1}%
\providecommand \href@noop [0]{\@secondoftwo}%
\providecommand \href [0]{\begingroup \@sanitize@url \@href}%
\providecommand \@href[1]{\@@startlink{#1}\@@href}%
\providecommand \@@href[1]{\endgroup#1\@@endlink}%
\providecommand \@sanitize@url [0]{\catcode `\\12\catcode `\$12\catcode `\&12\catcode `\#12\catcode `\^12\catcode `\_12\catcode `\%12\relax}%
\providecommand \@@startlink[1]{}%
\providecommand \@@endlink[0]{}%
\providecommand \url  [0]{\begingroup\@sanitize@url \@url }%
\providecommand \@url [1]{\endgroup\@href {#1}{\urlprefix }}%
\providecommand \urlprefix  [0]{URL }%
\providecommand \Eprint [0]{\href }%
\providecommand \doibase [0]{https://doi.org/}%
\providecommand \selectlanguage [0]{\@gobble}%
\providecommand \bibinfo  [0]{\@secondoftwo}%
\providecommand \bibfield  [0]{\@secondoftwo}%
\providecommand \translation [1]{[#1]}%
\providecommand \BibitemOpen [0]{}%
\providecommand \bibitemStop [0]{}%
\providecommand \bibitemNoStop [0]{.\EOS\space}%
\providecommand \EOS [0]{\spacefactor3000\relax}%
\providecommand \BibitemShut  [1]{\csname bibitem#1\endcsname}%
\let\auto@bib@innerbib\@empty
\bibitem [{\citenamefont {Calabrese}(2020)}]{10.21468/SciPostPhysLectNotes.20}%
  \BibitemOpen
  \bibfield  {author} {\bibinfo {author} {\bibfnamefont {P.}~\bibnamefont {Calabrese}},\ }\bibfield  {title} {\bibinfo {title} {{Entanglement spreading in non-equilibrium integrable systems}},\ }\href {https://doi.org/10.21468/SciPostPhysLectNotes.20} {\bibfield  {journal} {\bibinfo  {journal} {SciPost Phys. Lect. Notes}\ ,\ \bibinfo {pages} {20}} (\bibinfo {year} {2020})}\BibitemShut {NoStop}%
\bibitem [{\citenamefont {Osterloh}\ \emph {et~al.}(2002)\citenamefont {Osterloh}, \citenamefont {Amico}, \citenamefont {Falci},\ and\ \citenamefont {Fazio}}]{osterloh2002}%
  \BibitemOpen
  \bibfield  {author} {\bibinfo {author} {\bibfnamefont {A.}~\bibnamefont {Osterloh}}, \bibinfo {author} {\bibfnamefont {L.}~\bibnamefont {Amico}}, \bibinfo {author} {\bibfnamefont {G.}~\bibnamefont {Falci}},\ and\ \bibinfo {author} {\bibfnamefont {R.}~\bibnamefont {Fazio}},\ }\bibfield  {title} {\bibinfo {title} {Scaling of entanglement close to a quantum phase transition},\ }\href {https://doi.org/10.1038/416608a} {\bibfield  {journal} {\bibinfo  {journal} {Nature}\ }\textbf {\bibinfo {volume} {416}},\ \bibinfo {pages} {608} (\bibinfo {year} {2002})}\BibitemShut {NoStop}%
\bibitem [{\citenamefont {Baggioli}\ \emph {et~al.}(2023)\citenamefont {Baggioli}, \citenamefont {Liu},\ and\ \citenamefont {Wu}}]{Baggioli_2023}%
  \BibitemOpen
  \bibfield  {author} {\bibinfo {author} {\bibfnamefont {M.}~\bibnamefont {Baggioli}}, \bibinfo {author} {\bibfnamefont {Y.}~\bibnamefont {Liu}},\ and\ \bibinfo {author} {\bibfnamefont {X.-M.}\ \bibnamefont {Wu}},\ }\bibfield  {title} {\bibinfo {title} {Entanglement entropy as an order parameter for strongly coupled nodal line semimetals},\ }\href {https://doi.org/10.1007/JHEP05(2023)221} {\bibfield  {journal} {\bibinfo  {journal} {J. High Energ. Phys.}\ }\textbf {\bibinfo {volume} {2023}},\ \bibinfo {pages} {221 (2023)}}\BibitemShut {NoStop}%
\bibitem [{\citenamefont {Amico}\ \emph {et~al.}(2008)\citenamefont {Amico}, \citenamefont {Fazio}, \citenamefont {Osterloh},\ and\ \citenamefont {Vedral}}]{RevModPhys.80.517}%
  \BibitemOpen
  \bibfield  {author} {\bibinfo {author} {\bibfnamefont {L.}~\bibnamefont {Amico}}, \bibinfo {author} {\bibfnamefont {R.}~\bibnamefont {Fazio}}, \bibinfo {author} {\bibfnamefont {A.}~\bibnamefont {Osterloh}},\ and\ \bibinfo {author} {\bibfnamefont {V.}~\bibnamefont {Vedral}},\ }\bibfield  {title} {\bibinfo {title} {Entanglement in many-body systems},\ }\href {https://doi.org/10.1103/RevModPhys.80.517} {\bibfield  {journal} {\bibinfo  {journal} {Rev. Mod. Phys.}\ }\textbf {\bibinfo {volume} {80}},\ \bibinfo {pages} {517} (\bibinfo {year} {2008})}\BibitemShut {NoStop}%
\bibitem [{\citenamefont {Vidal}\ \emph {et~al.}(2003)\citenamefont {Vidal}, \citenamefont {Latorre}, \citenamefont {Rico},\ and\ \citenamefont {Kitaev}}]{PhysRevLett.90.227902}%
  \BibitemOpen
  \bibfield  {author} {\bibinfo {author} {\bibfnamefont {G.}~\bibnamefont {Vidal}}, \bibinfo {author} {\bibfnamefont {J.~I.}\ \bibnamefont {Latorre}}, \bibinfo {author} {\bibfnamefont {E.}~\bibnamefont {Rico}},\ and\ \bibinfo {author} {\bibfnamefont {A.}~\bibnamefont {Kitaev}},\ }\bibfield  {title} {\bibinfo {title} {Entanglement in quantum critical phenomena},\ }\href {https://doi.org/10.1103/PhysRevLett.90.227902} {\bibfield  {journal} {\bibinfo  {journal} {Phys. Rev. Lett.}\ }\textbf {\bibinfo {volume} {90}},\ \bibinfo {pages} {227902} (\bibinfo {year} {2003})}\BibitemShut {NoStop}%
\bibitem [{\citenamefont {Srednicki}(1994)}]{PhysRevE.50.888}%
  \BibitemOpen
  \bibfield  {author} {\bibinfo {author} {\bibfnamefont {M.}~\bibnamefont {Srednicki}},\ }\bibfield  {title} {\bibinfo {title} {Chaos and quantum thermalization},\ }\href {https://doi.org/10.1103/PhysRevE.50.888} {\bibfield  {journal} {\bibinfo  {journal} {Phys. Rev. E}\ }\textbf {\bibinfo {volume} {50}},\ \bibinfo {pages} {888} (\bibinfo {year} {1994})}\BibitemShut {NoStop}%
\bibitem [{\citenamefont {Rigol}\ \emph {et~al.}(2008)\citenamefont {Rigol}, \citenamefont {Dunjko},\ and\ \citenamefont {Olshanii}}]{Rigol_2008}%
  \BibitemOpen
  \bibfield  {author} {\bibinfo {author} {\bibfnamefont {M.}~\bibnamefont {Rigol}}, \bibinfo {author} {\bibfnamefont {V.}~\bibnamefont {Dunjko}},\ and\ \bibinfo {author} {\bibfnamefont {M.}~\bibnamefont {Olshanii}},\ }\bibfield  {title} {\bibinfo {title} {Thermalization and its mechanism for generic isolated quantum systems},\ }\href {https://doi.org/10.1038/nature06838} {\bibfield  {journal} {\bibinfo  {journal} {Nature}\ }\textbf {\bibinfo {volume} {452}},\ \bibinfo {pages} {854} (\bibinfo {year} {2008})}\BibitemShut {NoStop}%
\bibitem [{\citenamefont {Mallayya}\ \emph {et~al.}(2019)\citenamefont {Mallayya}, \citenamefont {Rigol},\ and\ \citenamefont {De~Roeck}}]{PhysRevX.9.021027}%
  \BibitemOpen
  \bibfield  {author} {\bibinfo {author} {\bibfnamefont {K.}~\bibnamefont {Mallayya}}, \bibinfo {author} {\bibfnamefont {M.}~\bibnamefont {Rigol}},\ and\ \bibinfo {author} {\bibfnamefont {W.}~\bibnamefont {De~Roeck}},\ }\bibfield  {title} {\bibinfo {title} {Prethermalization and thermalization in isolated quantum systems},\ }\href {https://doi.org/10.1103/PhysRevX.9.021027} {\bibfield  {journal} {\bibinfo  {journal} {Phys. Rev. X}\ }\textbf {\bibinfo {volume} {9}},\ \bibinfo {pages} {021027} (\bibinfo {year} {2019})}\BibitemShut {NoStop}%
\bibitem [{\citenamefont {Nahum}\ \emph {et~al.}(2017)\citenamefont {Nahum}, \citenamefont {Ruhman}, \citenamefont {Vijay},\ and\ \citenamefont {Haah}}]{PhysRevX.7.031016}%
  \BibitemOpen
  \bibfield  {author} {\bibinfo {author} {\bibfnamefont {A.}~\bibnamefont {Nahum}}, \bibinfo {author} {\bibfnamefont {J.}~\bibnamefont {Ruhman}}, \bibinfo {author} {\bibfnamefont {S.}~\bibnamefont {Vijay}},\ and\ \bibinfo {author} {\bibfnamefont {J.}~\bibnamefont {Haah}},\ }\bibfield  {title} {\bibinfo {title} {Quantum entanglement growth under random unitary dynamics},\ }\href {https://doi.org/10.1103/PhysRevX.7.031016} {\bibfield  {journal} {\bibinfo  {journal} {Phys. Rev. X}\ }\textbf {\bibinfo {volume} {7}},\ \bibinfo {pages} {031016} (\bibinfo {year} {2017})}\BibitemShut {NoStop}%
\bibitem [{\citenamefont {Fisher}\ \emph {et~al.}(2023)\citenamefont {Fisher}, \citenamefont {Khemani}, \citenamefont {Nahum},\ and\ \citenamefont {Vijay}}]{Fisher_2023}%
  \BibitemOpen
  \bibfield  {author} {\bibinfo {author} {\bibfnamefont {M.~P.}\ \bibnamefont {Fisher}}, \bibinfo {author} {\bibfnamefont {V.}~\bibnamefont {Khemani}}, \bibinfo {author} {\bibfnamefont {A.}~\bibnamefont {Nahum}},\ and\ \bibinfo {author} {\bibfnamefont {S.}~\bibnamefont {Vijay}},\ }\bibfield  {title} {\bibinfo {title} {Random quantum circuits},\ }\href {https://doi.org/10.1146/annurev-conmatphys-031720-030658} {\bibfield  {journal} {\bibinfo  {journal} {Annu. Rev. Condens. Matter Phys.}\ }\textbf {\bibinfo {volume} {14}},\ \bibinfo {pages} {335–379} (\bibinfo {year} {2023})}\BibitemShut {NoStop}%
\bibitem [{\citenamefont {Weinstein}\ \emph {et~al.}(2023)\citenamefont {Weinstein}, \citenamefont {Kelly}, \citenamefont {Marino},\ and\ \citenamefont {Altman}}]{Weinstein_2023}%
  \BibitemOpen
  \bibfield  {author} {\bibinfo {author} {\bibfnamefont {Z.}~\bibnamefont {Weinstein}}, \bibinfo {author} {\bibfnamefont {S.~P.}\ \bibnamefont {Kelly}}, \bibinfo {author} {\bibfnamefont {J.}~\bibnamefont {Marino}},\ and\ \bibinfo {author} {\bibfnamefont {E.}~\bibnamefont {Altman}},\ }\bibfield  {title} {\bibinfo {title} {Scrambling transition in a radiative random unitary circuit},\ }\href {https://doi.org/10.1103/physrevlett.131.220404} {\bibfield  {journal} {\bibinfo  {journal} {Phys. Rev. Lett.}\ }\textbf {\bibinfo {volume} {131}},\ \bibinfo {pages} {220404} (\bibinfo {year} {2023})}\BibitemShut {NoStop}%
\bibitem [{\citenamefont {Cheng}\ and\ \citenamefont {Ippoliti}(2023)}]{Cheng_2023}%
  \BibitemOpen
  \bibfield  {author} {\bibinfo {author} {\bibfnamefont {Z.}~\bibnamefont {Cheng}}\ and\ \bibinfo {author} {\bibfnamefont {M.}~\bibnamefont {Ippoliti}},\ }\bibfield  {title} {\bibinfo {title} {Efficient sampling of noisy shallow circuits via monitored unraveling},\ }\bibfield  {journal} {\bibinfo  {journal} {PRX Quantum}\ }\textbf {\bibinfo {volume} {4}},\ \href {https://doi.org/10.1103/prxquantum.4.040326} {10.1103/prxquantum.4.040326} (\bibinfo {year} {2023})\BibitemShut {NoStop}%
\bibitem [{\citenamefont {Chen}\ \emph {et~al.}(2023)\citenamefont {Chen}, \citenamefont {Bao},\ and\ \citenamefont {Choi}}]{chen2023}%
  \BibitemOpen
  \bibfield  {author} {\bibinfo {author} {\bibfnamefont {Z.}~\bibnamefont {Chen}}, \bibinfo {author} {\bibfnamefont {Y.}~\bibnamefont {Bao}},\ and\ \bibinfo {author} {\bibfnamefont {S.}~\bibnamefont {Choi}},\ }\href {https://arxiv.org/abs/2306.17161} {\bibinfo {title} {Optimized trajectory unraveling for classical simulation of noisy quantum dynamics}} (\bibinfo {year} {2023}),\ \Eprint {https://arxiv.org/abs/2306.17161} {arXiv:2306.17161 [quant-ph]} \BibitemShut {NoStop}%
\bibitem [{\citenamefont {Kolodrubetz}(2023)}]{PhysRevB.107.L140301}%
  \BibitemOpen
  \bibfield  {author} {\bibinfo {author} {\bibfnamefont {M.}~\bibnamefont {Kolodrubetz}},\ }\bibfield  {title} {\bibinfo {title} {Optimality of lindblad unfolding in measurement phase transitions},\ }\href {https://doi.org/10.1103/PhysRevB.107.L140301} {\bibfield  {journal} {\bibinfo  {journal} {Phys. Rev. B}\ }\textbf {\bibinfo {volume} {107}},\ \bibinfo {pages} {L140301} (\bibinfo {year} {2023})}\BibitemShut {NoStop}%
\bibitem [{\citenamefont {Calabrese}\ and\ \citenamefont {Cardy}(2005)}]{Calabrese_2005}%
  \BibitemOpen
  \bibfield  {author} {\bibinfo {author} {\bibfnamefont {P.}~\bibnamefont {Calabrese}}\ and\ \bibinfo {author} {\bibfnamefont {J.}~\bibnamefont {Cardy}},\ }\bibfield  {title} {\bibinfo {title} {Evolution of entanglement entropy in one-dimensional systems},\ }\href {https://doi.org/10.1088/1742-5468/2005/04/P04010} {\bibfield  {journal} {\bibinfo  {journal} {J. Stat. Mech.: Theory Exp.}\ }\textbf {\bibinfo {volume} {2005}},\ \bibinfo {pages} {P04010 (2005)}}\BibitemShut {NoStop}%
\bibitem [{\citenamefont {Skinner}\ \emph {et~al.}(2019)\citenamefont {Skinner}, \citenamefont {Ruhman},\ and\ \citenamefont {Nahum}}]{PhysRevX.9.031009}%
  \BibitemOpen
  \bibfield  {author} {\bibinfo {author} {\bibfnamefont {B.}~\bibnamefont {Skinner}}, \bibinfo {author} {\bibfnamefont {J.}~\bibnamefont {Ruhman}},\ and\ \bibinfo {author} {\bibfnamefont {A.}~\bibnamefont {Nahum}},\ }\bibfield  {title} {\bibinfo {title} {Measurement-induced phase transitions in the dynamics of entanglement},\ }\href {https://doi.org/10.1103/PhysRevX.9.031009} {\bibfield  {journal} {\bibinfo  {journal} {Phys. Rev. X}\ }\textbf {\bibinfo {volume} {9}},\ \bibinfo {pages} {031009} (\bibinfo {year} {2019})}\BibitemShut {NoStop}%
\bibitem [{\citenamefont {Li}\ \emph {et~al.}(2018)\citenamefont {Li}, \citenamefont {Chen},\ and\ \citenamefont {Fisher}}]{PhysRevB.98.205136}%
  \BibitemOpen
  \bibfield  {author} {\bibinfo {author} {\bibfnamefont {Y.}~\bibnamefont {Li}}, \bibinfo {author} {\bibfnamefont {X.}~\bibnamefont {Chen}},\ and\ \bibinfo {author} {\bibfnamefont {M.~P.~A.}\ \bibnamefont {Fisher}},\ }\bibfield  {title} {\bibinfo {title} {Quantum zeno effect and the many-body entanglement transition},\ }\href {https://doi.org/10.1103/PhysRevB.98.205136} {\bibfield  {journal} {\bibinfo  {journal} {Phys. Rev. B}\ }\textbf {\bibinfo {volume} {98}},\ \bibinfo {pages} {205136} (\bibinfo {year} {2018})}\BibitemShut {NoStop}%
\bibitem [{\citenamefont {Chan}\ \emph {et~al.}(2019)\citenamefont {Chan}, \citenamefont {Nandkishore}, \citenamefont {Pretko},\ and\ \citenamefont {Smith}}]{PhysRevB.99.224307}%
  \BibitemOpen
  \bibfield  {author} {\bibinfo {author} {\bibfnamefont {A.}~\bibnamefont {Chan}}, \bibinfo {author} {\bibfnamefont {R.~M.}\ \bibnamefont {Nandkishore}}, \bibinfo {author} {\bibfnamefont {M.}~\bibnamefont {Pretko}},\ and\ \bibinfo {author} {\bibfnamefont {G.}~\bibnamefont {Smith}},\ }\bibfield  {title} {\bibinfo {title} {Unitary-projective entanglement dynamics},\ }\href {https://doi.org/10.1103/PhysRevB.99.224307} {\bibfield  {journal} {\bibinfo  {journal} {Phys. Rev. B}\ }\textbf {\bibinfo {volume} {99}},\ \bibinfo {pages} {224307} (\bibinfo {year} {2019})}\BibitemShut {NoStop}%
\bibitem [{\citenamefont {Choi}\ \emph {et~al.}(2020)\citenamefont {Choi}, \citenamefont {Bao}, \citenamefont {Qi},\ and\ \citenamefont {Altman}}]{PhysRevLett.125.030505}%
  \BibitemOpen
  \bibfield  {author} {\bibinfo {author} {\bibfnamefont {S.}~\bibnamefont {Choi}}, \bibinfo {author} {\bibfnamefont {Y.}~\bibnamefont {Bao}}, \bibinfo {author} {\bibfnamefont {X.-L.}\ \bibnamefont {Qi}},\ and\ \bibinfo {author} {\bibfnamefont {E.}~\bibnamefont {Altman}},\ }\bibfield  {title} {\bibinfo {title} {Quantum error correction in scrambling dynamics and measurement-induced phase transition},\ }\href {https://doi.org/10.1103/PhysRevLett.125.030505} {\bibfield  {journal} {\bibinfo  {journal} {Phys. Rev. Lett.}\ }\textbf {\bibinfo {volume} {125}},\ \bibinfo {pages} {030505} (\bibinfo {year} {2020})}\BibitemShut {NoStop}%
\bibitem [{\citenamefont {Jian}\ \emph {et~al.}(2020)\citenamefont {Jian}, \citenamefont {You}, \citenamefont {Vasseur},\ and\ \citenamefont {Ludwig}}]{PhysRevB.101.104302}%
  \BibitemOpen
  \bibfield  {author} {\bibinfo {author} {\bibfnamefont {C.-M.}\ \bibnamefont {Jian}}, \bibinfo {author} {\bibfnamefont {Y.-Z.}\ \bibnamefont {You}}, \bibinfo {author} {\bibfnamefont {R.}~\bibnamefont {Vasseur}},\ and\ \bibinfo {author} {\bibfnamefont {A.~W.~W.}\ \bibnamefont {Ludwig}},\ }\bibfield  {title} {\bibinfo {title} {Measurement-induced criticality in random quantum circuits},\ }\href {https://doi.org/10.1103/PhysRevB.101.104302} {\bibfield  {journal} {\bibinfo  {journal} {Phys. Rev. B}\ }\textbf {\bibinfo {volume} {101}},\ \bibinfo {pages} {104302} (\bibinfo {year} {2020})}\BibitemShut {NoStop}%
\bibitem [{\citenamefont {Zhang}\ \emph {et~al.}(2020)\citenamefont {Zhang}, \citenamefont {Reyes}, \citenamefont {Kourtis}, \citenamefont {Chamon}, \citenamefont {Mucciolo},\ and\ \citenamefont {Ruckenstein}}]{PhysRevB.101.235104}%
  \BibitemOpen
  \bibfield  {author} {\bibinfo {author} {\bibfnamefont {L.}~\bibnamefont {Zhang}}, \bibinfo {author} {\bibfnamefont {J.~A.}\ \bibnamefont {Reyes}}, \bibinfo {author} {\bibfnamefont {S.}~\bibnamefont {Kourtis}}, \bibinfo {author} {\bibfnamefont {C.}~\bibnamefont {Chamon}}, \bibinfo {author} {\bibfnamefont {E.~R.}\ \bibnamefont {Mucciolo}},\ and\ \bibinfo {author} {\bibfnamefont {A.~E.}\ \bibnamefont {Ruckenstein}},\ }\bibfield  {title} {\bibinfo {title} {Nonuniversal entanglement level statistics in projection-driven quantum circuits},\ }\href {https://doi.org/10.1103/PhysRevB.101.235104} {\bibfield  {journal} {\bibinfo  {journal} {Phys. Rev. B}\ }\textbf {\bibinfo {volume} {101}},\ \bibinfo {pages} {235104} (\bibinfo {year} {2020})}\BibitemShut {NoStop}%
\bibitem [{\citenamefont {Kelly}\ and\ \citenamefont {Marino}(2024)}]{kelly2024}%
  \BibitemOpen
  \bibfield  {author} {\bibinfo {author} {\bibfnamefont {S.~P.}\ \bibnamefont {Kelly}}\ and\ \bibinfo {author} {\bibfnamefont {J.}~\bibnamefont {Marino}},\ }\href@noop {} {\bibinfo {title} {Generalizing measurement-induced phase transitions to information exchange symmetry breaking}} (\bibinfo {year} {2024}),\ \Eprint {https://arxiv.org/abs/2402.13271} {arXiv:2402.13271 [quant-ph]} \BibitemShut {NoStop}%
\bibitem [{\citenamefont {Jacobs}\ and\ \citenamefont {Steck}(2006)}]{Jacobs_2006}%
  \BibitemOpen
  \bibfield  {author} {\bibinfo {author} {\bibfnamefont {K.}~\bibnamefont {Jacobs}}\ and\ \bibinfo {author} {\bibfnamefont {D.~A.}\ \bibnamefont {Steck}},\ }\bibfield  {title} {\bibinfo {title} {A straightforward introduction to continuous quantum measurement},\ }\href {https://doi.org/10.1080/00107510601101934} {\bibfield  {journal} {\bibinfo  {journal} {Contemp. Phys.}\ }\textbf {\bibinfo {volume} {47}},\ \bibinfo {pages} {279} (\bibinfo {year} {2006})}\BibitemShut {NoStop}%
\bibitem [{\citenamefont {Li}\ \emph {et~al.}(2019)\citenamefont {Li}, \citenamefont {Chen},\ and\ \citenamefont {Fisher}}]{PhysRevB.100.134306}%
  \BibitemOpen
  \bibfield  {author} {\bibinfo {author} {\bibfnamefont {Y.}~\bibnamefont {Li}}, \bibinfo {author} {\bibfnamefont {X.}~\bibnamefont {Chen}},\ and\ \bibinfo {author} {\bibfnamefont {M.~P.~A.}\ \bibnamefont {Fisher}},\ }\bibfield  {title} {\bibinfo {title} {Measurement-driven entanglement transition in hybrid quantum circuits},\ }\href {https://doi.org/10.1103/PhysRevB.100.134306} {\bibfield  {journal} {\bibinfo  {journal} {Phys. Rev. B}\ }\textbf {\bibinfo {volume} {100}},\ \bibinfo {pages} {134306} (\bibinfo {year} {2019})}\BibitemShut {NoStop}%
\bibitem [{\citenamefont {Cao}\ \emph {et~al.}(2019)\citenamefont {Cao}, \citenamefont {Tilloy},\ and\ \citenamefont {Luca}}]{10.21468/SciPostPhys.7.2.024}%
  \BibitemOpen
  \bibfield  {author} {\bibinfo {author} {\bibfnamefont {X.}~\bibnamefont {Cao}}, \bibinfo {author} {\bibfnamefont {A.}~\bibnamefont {Tilloy}},\ and\ \bibinfo {author} {\bibfnamefont {A.~D.}\ \bibnamefont {Luca}},\ }\bibfield  {title} {\bibinfo {title} {{Entanglement in a fermion chain under continuous monitoring}},\ }\href {https://doi.org/10.21468/SciPostPhys.7.2.024} {\bibfield  {journal} {\bibinfo  {journal} {SciPost Phys.}\ }\textbf {\bibinfo {volume} {7}},\ \bibinfo {pages} {024} (\bibinfo {year} {2019})}\BibitemShut {NoStop}%
\bibitem [{\citenamefont {Gullans}\ and\ \citenamefont {Huse}(2020{\natexlab{a}})}]{PhysRevX.10.041020}%
  \BibitemOpen
  \bibfield  {author} {\bibinfo {author} {\bibfnamefont {M.~J.}\ \bibnamefont {Gullans}}\ and\ \bibinfo {author} {\bibfnamefont {D.~A.}\ \bibnamefont {Huse}},\ }\bibfield  {title} {\bibinfo {title} {Dynamical purification phase transition induced by quantum measurements},\ }\href {https://doi.org/10.1103/PhysRevX.10.041020} {\bibfield  {journal} {\bibinfo  {journal} {Phys. Rev. X}\ }\textbf {\bibinfo {volume} {10}},\ \bibinfo {pages} {041020} (\bibinfo {year} {2020}{\natexlab{a}})}\BibitemShut {NoStop}%
\bibitem [{\citenamefont {Alberton}\ \emph {et~al.}(2021)\citenamefont {Alberton}, \citenamefont {Buchhold},\ and\ \citenamefont {Diehl}}]{Alberton_2021}%
  \BibitemOpen
  \bibfield  {author} {\bibinfo {author} {\bibfnamefont {O.}~\bibnamefont {Alberton}}, \bibinfo {author} {\bibfnamefont {M.}~\bibnamefont {Buchhold}},\ and\ \bibinfo {author} {\bibfnamefont {S.}~\bibnamefont {Diehl}},\ }\bibfield  {title} {\bibinfo {title} {Entanglement transition in a monitored free-fermion chain: From extended criticality to area law},\ }\href {https://doi.org/10.1103/PhysRevLett.126.170602} {\bibfield  {journal} {\bibinfo  {journal} {Phys. Rev. Lett.}\ }\textbf {\bibinfo {volume} {126}},\ \bibinfo {pages} {170602} (\bibinfo {year} {2021})}\BibitemShut {NoStop}%
\bibitem [{\citenamefont {Turkeshi}\ \emph {et~al.}(2021)\citenamefont {Turkeshi}, \citenamefont {Biella}, \citenamefont {Fazio}, \citenamefont {Dalmonte},\ and\ \citenamefont {Schir\'o}}]{Turkeshi_2021}%
  \BibitemOpen
  \bibfield  {author} {\bibinfo {author} {\bibfnamefont {X.}~\bibnamefont {Turkeshi}}, \bibinfo {author} {\bibfnamefont {A.}~\bibnamefont {Biella}}, \bibinfo {author} {\bibfnamefont {R.}~\bibnamefont {Fazio}}, \bibinfo {author} {\bibfnamefont {M.}~\bibnamefont {Dalmonte}},\ and\ \bibinfo {author} {\bibfnamefont {M.}~\bibnamefont {Schir\'o}},\ }\bibfield  {title} {\bibinfo {title} {Measurement-induced entanglement transitions in the quantum ising chain: From infinite to zero clicks},\ }\href {https://doi.org/10.1103/PhysRevB.103.224210} {\bibfield  {journal} {\bibinfo  {journal} {Phys. Rev. B}\ }\textbf {\bibinfo {volume} {103}},\ \bibinfo {pages} {224210} (\bibinfo {year} {2021})}\BibitemShut {NoStop}%
\bibitem [{\citenamefont {Turkeshi}\ \emph {et~al.}(2022)\citenamefont {Turkeshi}, \citenamefont {Dalmonte}, \citenamefont {Fazio},\ and\ \citenamefont {Schir\`o}}]{Turkeshi_2022}%
  \BibitemOpen
  \bibfield  {author} {\bibinfo {author} {\bibfnamefont {X.}~\bibnamefont {Turkeshi}}, \bibinfo {author} {\bibfnamefont {M.}~\bibnamefont {Dalmonte}}, \bibinfo {author} {\bibfnamefont {R.}~\bibnamefont {Fazio}},\ and\ \bibinfo {author} {\bibfnamefont {M.}~\bibnamefont {Schir\`o}},\ }\bibfield  {title} {\bibinfo {title} {Entanglement transitions from stochastic resetting of non-hermitian quasiparticles},\ }\href {https://doi.org/10.1103/PhysRevB.105.L241114} {\bibfield  {journal} {\bibinfo  {journal} {Phys. Rev. B}\ }\textbf {\bibinfo {volume} {105}},\ \bibinfo {pages} {L241114} (\bibinfo {year} {2022})}\BibitemShut {NoStop}%
\bibitem [{\citenamefont {Coppola}\ \emph {et~al.}(2022)\citenamefont {Coppola}, \citenamefont {Tirrito}, \citenamefont {Karevski},\ and\ \citenamefont {Collura}}]{Coppola_2022}%
  \BibitemOpen
  \bibfield  {author} {\bibinfo {author} {\bibfnamefont {M.}~\bibnamefont {Coppola}}, \bibinfo {author} {\bibfnamefont {E.}~\bibnamefont {Tirrito}}, \bibinfo {author} {\bibfnamefont {D.}~\bibnamefont {Karevski}},\ and\ \bibinfo {author} {\bibfnamefont {M.}~\bibnamefont {Collura}},\ }\bibfield  {title} {\bibinfo {title} {Growth of entanglement entropy under local projective measurements},\ }\href {https://doi.org/10.1103/PhysRevB.105.094303} {\bibfield  {journal} {\bibinfo  {journal} {Phys. Rev. B}\ }\textbf {\bibinfo {volume} {105}},\ \bibinfo {pages} {094303} (\bibinfo {year} {2022})}\BibitemShut {NoStop}%
\bibitem [{\citenamefont {Carollo}\ and\ \citenamefont {Alba}(2022)}]{PhysRevB.106.L220304}%
  \BibitemOpen
  \bibfield  {author} {\bibinfo {author} {\bibfnamefont {F.}~\bibnamefont {Carollo}}\ and\ \bibinfo {author} {\bibfnamefont {V.}~\bibnamefont {Alba}},\ }\bibfield  {title} {\bibinfo {title} {Entangled multiplets and spreading of quantum correlations in a continuously monitored tight-binding chain},\ }\href {https://doi.org/10.1103/PhysRevB.106.L220304} {\bibfield  {journal} {\bibinfo  {journal} {Phys. Rev. B}\ }\textbf {\bibinfo {volume} {106}},\ \bibinfo {pages} {L220304} (\bibinfo {year} {2022})}\BibitemShut {NoStop}%
\bibitem [{\citenamefont {Vovk}\ and\ \citenamefont {Pichler}(2022)}]{PhysRevLett.128.243601}%
  \BibitemOpen
  \bibfield  {author} {\bibinfo {author} {\bibfnamefont {T.}~\bibnamefont {Vovk}}\ and\ \bibinfo {author} {\bibfnamefont {H.}~\bibnamefont {Pichler}},\ }\bibfield  {title} {\bibinfo {title} {Entanglement-optimal trajectories of many-body quantum markov processes},\ }\href {https://doi.org/10.1103/PhysRevLett.128.243601} {\bibfield  {journal} {\bibinfo  {journal} {Phys. Rev. Lett.}\ }\textbf {\bibinfo {volume} {128}},\ \bibinfo {pages} {243601} (\bibinfo {year} {2022})}\BibitemShut {NoStop}%
\bibitem [{\citenamefont {Gal}\ \emph {et~al.}(2023)\citenamefont {Gal}, \citenamefont {Turkeshi},\ and\ \citenamefont {Schirò}}]{Le_Gal_2023}%
  \BibitemOpen
  \bibfield  {author} {\bibinfo {author} {\bibfnamefont {Y.~L.}\ \bibnamefont {Gal}}, \bibinfo {author} {\bibfnamefont {X.}~\bibnamefont {Turkeshi}},\ and\ \bibinfo {author} {\bibfnamefont {M.}~\bibnamefont {Schirò}},\ }\bibfield  {title} {\bibinfo {title} {{Volume-to-area law entanglement transition in a non-Hermitian free fermionic chain}},\ }\href {https://doi.org/10.21468/SciPostPhys.14.5.138} {\bibfield  {journal} {\bibinfo  {journal} {SciPost Phys.}\ }\textbf {\bibinfo {volume} {14}},\ \bibinfo {pages} {138} (\bibinfo {year} {2023})}\BibitemShut {NoStop}%
\bibitem [{\citenamefont {Granet}\ \emph {et~al.}(2023)\citenamefont {Granet}, \citenamefont {Zhang},\ and\ \citenamefont {Dreyer}}]{PhysRevLett.130.230401}%
  \BibitemOpen
  \bibfield  {author} {\bibinfo {author} {\bibfnamefont {E.}~\bibnamefont {Granet}}, \bibinfo {author} {\bibfnamefont {C.}~\bibnamefont {Zhang}},\ and\ \bibinfo {author} {\bibfnamefont {H.}~\bibnamefont {Dreyer}},\ }\bibfield  {title} {\bibinfo {title} {{Volume-Law to Area-Law Entanglement Transition in a Nonunitary Periodic Gaussian Circuit}},\ }\href {https://doi.org/10.1103/PhysRevLett.130.230401} {\bibfield  {journal} {\bibinfo  {journal} {Phys. Rev. Lett.}\ }\textbf {\bibinfo {volume} {130}},\ \bibinfo {pages} {230401} (\bibinfo {year} {2023})}\BibitemShut {NoStop}%
\bibitem [{\citenamefont {Ippoliti}\ \emph {et~al.}(2021)\citenamefont {Ippoliti}, \citenamefont {Gullans}, \citenamefont {Gopalakrishnan}, \citenamefont {Huse},\ and\ \citenamefont {Khemani}}]{Ippoliti_2021}%
  \BibitemOpen
  \bibfield  {author} {\bibinfo {author} {\bibfnamefont {M.}~\bibnamefont {Ippoliti}}, \bibinfo {author} {\bibfnamefont {M.~J.}\ \bibnamefont {Gullans}}, \bibinfo {author} {\bibfnamefont {S.}~\bibnamefont {Gopalakrishnan}}, \bibinfo {author} {\bibfnamefont {D.~A.}\ \bibnamefont {Huse}},\ and\ \bibinfo {author} {\bibfnamefont {V.}~\bibnamefont {Khemani}},\ }\bibfield  {title} {\bibinfo {title} {Entanglement phase transitions in measurement-only dynamics},\ }\href {https://doi.org/10.1103/PhysRevX.11.011030} {\bibfield  {journal} {\bibinfo  {journal} {Phys. Rev. X}\ }\textbf {\bibinfo {volume} {11}},\ \bibinfo {pages} {011030} (\bibinfo {year} {2021})}\BibitemShut {NoStop}%
\bibitem [{\citenamefont {Tang}\ and\ \citenamefont {Zhu}(2020)}]{PhysRevResearch.2.013022}%
  \BibitemOpen
  \bibfield  {author} {\bibinfo {author} {\bibfnamefont {Q.}~\bibnamefont {Tang}}\ and\ \bibinfo {author} {\bibfnamefont {W.}~\bibnamefont {Zhu}},\ }\bibfield  {title} {\bibinfo {title} {Measurement-induced phase transition: A case study in the nonintegrable model by density-matrix renormalization group calculations},\ }\href {https://doi.org/10.1103/PhysRevResearch.2.013022} {\bibfield  {journal} {\bibinfo  {journal} {Phys. Rev. Res.}\ }\textbf {\bibinfo {volume} {2}},\ \bibinfo {pages} {013022} (\bibinfo {year} {2020})}\BibitemShut {NoStop}%
\bibitem [{\citenamefont {Vovk}\ and\ \citenamefont {Pichler}(2024)}]{vovk2024quantum}%
  \BibitemOpen
  \bibfield  {author} {\bibinfo {author} {\bibfnamefont {T.}~\bibnamefont {Vovk}}\ and\ \bibinfo {author} {\bibfnamefont {H.}~\bibnamefont {Pichler}},\ }\href@noop {} {\bibinfo {title} {Quantum trajectory entanglement in various unravelings of markovian dynamics}} (\bibinfo {year} {2024}),\ \Eprint {https://arxiv.org/abs/2404.12167} {arXiv:2404.12167 [quant-ph]} \BibitemShut {NoStop}%
\bibitem [{\citenamefont {Zabalo}\ \emph {et~al.}(2020)\citenamefont {Zabalo}, \citenamefont {Gullans}, \citenamefont {Wilson}, \citenamefont {Gopalakrishnan}, \citenamefont {Huse},\ and\ \citenamefont {Pixley}}]{PhysRevB.101.060301}%
  \BibitemOpen
  \bibfield  {author} {\bibinfo {author} {\bibfnamefont {A.}~\bibnamefont {Zabalo}}, \bibinfo {author} {\bibfnamefont {M.~J.}\ \bibnamefont {Gullans}}, \bibinfo {author} {\bibfnamefont {J.~H.}\ \bibnamefont {Wilson}}, \bibinfo {author} {\bibfnamefont {S.}~\bibnamefont {Gopalakrishnan}}, \bibinfo {author} {\bibfnamefont {D.~A.}\ \bibnamefont {Huse}},\ and\ \bibinfo {author} {\bibfnamefont {J.~H.}\ \bibnamefont {Pixley}},\ }\bibfield  {title} {\bibinfo {title} {Critical properties of the measurement-induced transition in random quantum circuits},\ }\href {https://doi.org/10.1103/PhysRevB.101.060301} {\bibfield  {journal} {\bibinfo  {journal} {Phys. Rev. B}\ }\textbf {\bibinfo {volume} {101}},\ \bibinfo {pages} {060301} (\bibinfo {year} {2020})}\BibitemShut {NoStop}%
\bibitem [{\citenamefont {Bao}\ \emph {et~al.}(2020)\citenamefont {Bao}, \citenamefont {Choi},\ and\ \citenamefont {Altman}}]{PhysRevB.101.104301}%
  \BibitemOpen
  \bibfield  {author} {\bibinfo {author} {\bibfnamefont {Y.}~\bibnamefont {Bao}}, \bibinfo {author} {\bibfnamefont {S.}~\bibnamefont {Choi}},\ and\ \bibinfo {author} {\bibfnamefont {E.}~\bibnamefont {Altman}},\ }\bibfield  {title} {\bibinfo {title} {Theory of the phase transition in random unitary circuits with measurements},\ }\href {https://doi.org/10.1103/PhysRevB.101.104301} {\bibfield  {journal} {\bibinfo  {journal} {Phys. Rev. B}\ }\textbf {\bibinfo {volume} {101}},\ \bibinfo {pages} {104301} (\bibinfo {year} {2020})}\BibitemShut {NoStop}%
\bibitem [{\citenamefont {Gullans}\ and\ \citenamefont {Huse}(2020{\natexlab{b}})}]{PhysRevLett.125.070606}%
  \BibitemOpen
  \bibfield  {author} {\bibinfo {author} {\bibfnamefont {M.~J.}\ \bibnamefont {Gullans}}\ and\ \bibinfo {author} {\bibfnamefont {D.~A.}\ \bibnamefont {Huse}},\ }\bibfield  {title} {\bibinfo {title} {Scalable probes of measurement-induced criticality},\ }\href {https://doi.org/10.1103/PhysRevLett.125.070606} {\bibfield  {journal} {\bibinfo  {journal} {Phys. Rev. Lett.}\ }\textbf {\bibinfo {volume} {125}},\ \bibinfo {pages} {070606} (\bibinfo {year} {2020}{\natexlab{b}})}\BibitemShut {NoStop}%
\bibitem [{\citenamefont {Nahum}\ and\ \citenamefont {Skinner}(2020)}]{PhysRevResearch.2.023288}%
  \BibitemOpen
  \bibfield  {author} {\bibinfo {author} {\bibfnamefont {A.}~\bibnamefont {Nahum}}\ and\ \bibinfo {author} {\bibfnamefont {B.}~\bibnamefont {Skinner}},\ }\bibfield  {title} {\bibinfo {title} {Entanglement and dynamics of diffusion-annihilation processes with majorana defects},\ }\href {https://doi.org/10.1103/PhysRevResearch.2.023288} {\bibfield  {journal} {\bibinfo  {journal} {Phys. Rev. Res.}\ }\textbf {\bibinfo {volume} {2}},\ \bibinfo {pages} {023288} (\bibinfo {year} {2020})}\BibitemShut {NoStop}%
\bibitem [{\citenamefont {Minato}\ \emph {et~al.}(2022)\citenamefont {Minato}, \citenamefont {Sugimoto}, \citenamefont {Kuwahara},\ and\ \citenamefont {Saito}}]{minato2022}%
  \BibitemOpen
  \bibfield  {author} {\bibinfo {author} {\bibfnamefont {T.}~\bibnamefont {Minato}}, \bibinfo {author} {\bibfnamefont {K.}~\bibnamefont {Sugimoto}}, \bibinfo {author} {\bibfnamefont {T.}~\bibnamefont {Kuwahara}},\ and\ \bibinfo {author} {\bibfnamefont {K.}~\bibnamefont {Saito}},\ }\bibfield  {title} {\bibinfo {title} {Fate of measurement-induced phase transition in long-range interactions},\ }\href {https://doi.org/10.1103/PhysRevLett.128.010603} {\bibfield  {journal} {\bibinfo  {journal} {Phys. Rev. Lett.}\ }\textbf {\bibinfo {volume} {128}},\ \bibinfo {pages} {010603} (\bibinfo {year} {2022})}\BibitemShut {NoStop}%
\bibitem [{\citenamefont {Wiseman}\ and\ \citenamefont {Diósi}(2001)}]{wiseman2001}%
  \BibitemOpen
  \bibfield  {author} {\bibinfo {author} {\bibfnamefont {H.}~\bibnamefont {Wiseman}}\ and\ \bibinfo {author} {\bibfnamefont {L.}~\bibnamefont {Diósi}},\ }\bibfield  {title} {\bibinfo {title} {Complete parameterization, and invariance, of diffusive quantum trajectories for markovian open systems},\ }\href {https://doi.org/https://doi.org/10.1016/S0301-0104(01)00296-8} {\bibfield  {journal} {\bibinfo  {journal} {Chem. Phys.}\ }\textbf {\bibinfo {volume} {268}},\ \bibinfo {pages} {91} (\bibinfo {year} {2001})}\BibitemShut {NoStop}%
\bibitem [{\citenamefont {Wiseman}\ and\ \citenamefont {Milburn}(2009)}]{wiseman2009}%
  \BibitemOpen
  \bibfield  {author} {\bibinfo {author} {\bibfnamefont {H.~M.}\ \bibnamefont {Wiseman}}\ and\ \bibinfo {author} {\bibfnamefont {G.~J.}\ \bibnamefont {Milburn}},\ }\href {https://doi.org/https://doi.org/10.1017/CBO9780511813948} {\emph {\bibinfo {title} {Quantum measurement and control}}}\ (\bibinfo  {publisher} {Cambridge university press},\ \bibinfo {year} {2009})\BibitemShut {NoStop}%
\bibitem [{\citenamefont {Genoni}\ \emph {et~al.}(2014)\citenamefont {Genoni}, \citenamefont {Mancini},\ and\ \citenamefont {Serafini}}]{genoni2014}%
  \BibitemOpen
  \bibfield  {author} {\bibinfo {author} {\bibfnamefont {M.~G.}\ \bibnamefont {Genoni}}, \bibinfo {author} {\bibfnamefont {S.}~\bibnamefont {Mancini}},\ and\ \bibinfo {author} {\bibfnamefont {A.}~\bibnamefont {Serafini}},\ }\bibfield  {title} {\bibinfo {title} {General-dyne unravelling of a thermal master equation},\ }\href {https://doi.org/10.1134/S1061920814030054} {\bibfield  {journal} {\bibinfo  {journal} {Russ. J. Math. Phys.}\ }\textbf {\bibinfo {volume} {21}},\ \bibinfo {pages} {329} (\bibinfo {year} {2014})}\BibitemShut {NoStop}%
\bibitem [{\citenamefont {Clarke}\ \emph {et~al.}(2023)\citenamefont {Clarke}, \citenamefont {Neveu}, \citenamefont {Khosla}, \citenamefont {Verhagen},\ and\ \citenamefont {Vanner}}]{clarke2023}%
  \BibitemOpen
  \bibfield  {author} {\bibinfo {author} {\bibfnamefont {J.}~\bibnamefont {Clarke}}, \bibinfo {author} {\bibfnamefont {P.}~\bibnamefont {Neveu}}, \bibinfo {author} {\bibfnamefont {K.~E.}\ \bibnamefont {Khosla}}, \bibinfo {author} {\bibfnamefont {E.}~\bibnamefont {Verhagen}},\ and\ \bibinfo {author} {\bibfnamefont {M.~R.}\ \bibnamefont {Vanner}},\ }\bibfield  {title} {\bibinfo {title} {Cavity quantum optomechanical nonlinearities and position measurement beyond the breakdown of the linearized approximation},\ }\href {https://doi.org/10.1103/PhysRevLett.131.053601} {\bibfield  {journal} {\bibinfo  {journal} {Phys. Rev. Lett.}\ }\textbf {\bibinfo {volume} {131}},\ \bibinfo {pages} {053601} (\bibinfo {year} {2023})}\BibitemShut {NoStop}%
\bibitem [{\citenamefont {Yang}\ \emph {et~al.}(2018)\citenamefont {Yang}, \citenamefont {Laflamme}, \citenamefont {Vasilyev}, \citenamefont {Baranov},\ and\ \citenamefont {Zoller}}]{PhysRevLett.120.133601}%
  \BibitemOpen
  \bibfield  {author} {\bibinfo {author} {\bibfnamefont {D.}~\bibnamefont {Yang}}, \bibinfo {author} {\bibfnamefont {C.}~\bibnamefont {Laflamme}}, \bibinfo {author} {\bibfnamefont {D.~V.}\ \bibnamefont {Vasilyev}}, \bibinfo {author} {\bibfnamefont {M.~A.}\ \bibnamefont {Baranov}},\ and\ \bibinfo {author} {\bibfnamefont {P.}~\bibnamefont {Zoller}},\ }\bibfield  {title} {\bibinfo {title} {Theory of a quantum scanning microscope for cold atoms},\ }\href {https://doi.org/10.1103/PhysRevLett.120.133601} {\bibfield  {journal} {\bibinfo  {journal} {Phys. Rev. Lett.}\ }\textbf {\bibinfo {volume} {120}},\ \bibinfo {pages} {133601} (\bibinfo {year} {2018})}\BibitemShut {NoStop}%
\bibitem [{\citenamefont {Young}\ \emph {et~al.}(2024)\citenamefont {Young}, \citenamefont {Gangardt},\ and\ \citenamefont {von Keyserlingk}}]{young2024}%
  \BibitemOpen
  \bibfield  {author} {\bibinfo {author} {\bibfnamefont {T.}~\bibnamefont {Young}}, \bibinfo {author} {\bibfnamefont {D.~M.}\ \bibnamefont {Gangardt}},\ and\ \bibinfo {author} {\bibfnamefont {C.}~\bibnamefont {von Keyserlingk}},\ }\href {https://arxiv.org/abs/2403.04022} {\bibinfo {title} {Diffusive entanglement growth in a monitored harmonic chain}} (\bibinfo {year} {2024}),\ \Eprint {https://arxiv.org/abs/2403.04022} {arXiv:2403.04022 [quant-ph]} \BibitemShut {NoStop}%
\bibitem [{\citenamefont {Piccitto}\ \emph {et~al.}(2022)\citenamefont {Piccitto}, \citenamefont {Russomanno},\ and\ \citenamefont {Rossini}}]{piccitto2022}%
  \BibitemOpen
  \bibfield  {author} {\bibinfo {author} {\bibfnamefont {G.}~\bibnamefont {Piccitto}}, \bibinfo {author} {\bibfnamefont {A.}~\bibnamefont {Russomanno}},\ and\ \bibinfo {author} {\bibfnamefont {D.}~\bibnamefont {Rossini}},\ }\bibfield  {title} {\bibinfo {title} {Entanglement transitions in the quantum ising chain: A comparison between different unravelings of the same lindbladian},\ }\href {https://doi.org/10.1103/PhysRevB.105.064305} {\bibfield  {journal} {\bibinfo  {journal} {Phys. Rev. B}\ }\textbf {\bibinfo {volume} {105}},\ \bibinfo {pages} {064305} (\bibinfo {year} {2022})}\BibitemShut {NoStop}%
\bibitem [{\citenamefont {Peschel}(2003)}]{Peschel_2003}%
  \BibitemOpen
  \bibfield  {author} {\bibinfo {author} {\bibfnamefont {I.}~\bibnamefont {Peschel}},\ }\bibfield  {title} {\bibinfo {title} {Calculation of reduced density matrices from correlation functions},\ }\href {https://doi.org/10.1088/0305-4470/36/14/101} {\bibfield  {journal} {\bibinfo  {journal} {J. Phys. A: Math. Gen.}\ }\textbf {\bibinfo {volume} {36}},\ \bibinfo {pages} {L205} (\bibinfo {year} {2003})}\BibitemShut {NoStop}%
\bibitem [{\citenamefont {Peschel}\ and\ \citenamefont {Eisler}(2009)}]{Peschel_2009}%
  \BibitemOpen
  \bibfield  {author} {\bibinfo {author} {\bibfnamefont {I.}~\bibnamefont {Peschel}}\ and\ \bibinfo {author} {\bibfnamefont {V.}~\bibnamefont {Eisler}},\ }\bibfield  {title} {\bibinfo {title} {Reduced density matrices and entanglement entropy in free lattice models},\ }\href {https://doi.org/10.1088/1751-8113/42/50/504003} {\bibfield  {journal} {\bibinfo  {journal} {J. Phys. A: Math. Theor.}\ }\textbf {\bibinfo {volume} {42}},\ \bibinfo {pages} {504003} (\bibinfo {year} {2009})}\BibitemShut {NoStop}%
\bibitem [{\citenamefont {Holzhey}\ \emph {et~al.}(1994)\citenamefont {Holzhey}, \citenamefont {Larsen},\ and\ \citenamefont {Wilczek}}]{Holzhey_1994}%
  \BibitemOpen
  \bibfield  {author} {\bibinfo {author} {\bibfnamefont {C.}~\bibnamefont {Holzhey}}, \bibinfo {author} {\bibfnamefont {F.}~\bibnamefont {Larsen}},\ and\ \bibinfo {author} {\bibfnamefont {F.}~\bibnamefont {Wilczek}},\ }\bibfield  {title} {\bibinfo {title} {Geometric and renormalized entropy in conformal field theory},\ }\href {https://doi.org/10.1016/0550-3213(94)90402-2} {\bibfield  {journal} {\bibinfo  {journal} {Nucl. Phys. B}\ }\textbf {\bibinfo {volume} {424}},\ \bibinfo {pages} {443–467} (\bibinfo {year} {1994})}\BibitemShut {NoStop}%
\bibitem [{\citenamefont {Calabrese}\ and\ \citenamefont {Cardy}(2009)}]{Calabrese_2009}%
  \BibitemOpen
  \bibfield  {author} {\bibinfo {author} {\bibfnamefont {P.}~\bibnamefont {Calabrese}}\ and\ \bibinfo {author} {\bibfnamefont {J.}~\bibnamefont {Cardy}},\ }\bibfield  {title} {\bibinfo {title} {Entanglement entropy and conformal field theory},\ }\href {https://doi.org/10.1088/1751-8113/42/50/504005} {\bibfield  {journal} {\bibinfo  {journal} {J. Phys. A: Math. Theor.}\ }\textbf {\bibinfo {volume} {42}},\ \bibinfo {pages} {504005} (\bibinfo {year} {2009})}\BibitemShut {NoStop}%
\bibitem [{\citenamefont {Kaplan}\ \emph {et~al.}(2009)\citenamefont {Kaplan}, \citenamefont {Lee}, \citenamefont {Son},\ and\ \citenamefont {Stephanov}}]{PhysRevD.80.125005}%
  \BibitemOpen
  \bibfield  {author} {\bibinfo {author} {\bibfnamefont {D.~B.}\ \bibnamefont {Kaplan}}, \bibinfo {author} {\bibfnamefont {J.-W.}\ \bibnamefont {Lee}}, \bibinfo {author} {\bibfnamefont {D.~T.}\ \bibnamefont {Son}},\ and\ \bibinfo {author} {\bibfnamefont {M.~A.}\ \bibnamefont {Stephanov}},\ }\bibfield  {title} {\bibinfo {title} {Conformality lost},\ }\href {https://doi.org/10.1103/PhysRevD.80.125005} {\bibfield  {journal} {\bibinfo  {journal} {Phys. Rev. D}\ }\textbf {\bibinfo {volume} {80}},\ \bibinfo {pages} {125005} (\bibinfo {year} {2009})}\BibitemShut {NoStop}%
\bibitem [{\citenamefont {Carrasquilla}\ and\ \citenamefont {Rigol}(2012)}]{PhysRevA.86.043629}%
  \BibitemOpen
  \bibfield  {author} {\bibinfo {author} {\bibfnamefont {J.}~\bibnamefont {Carrasquilla}}\ and\ \bibinfo {author} {\bibfnamefont {M.}~\bibnamefont {Rigol}},\ }\bibfield  {title} {\bibinfo {title} {Superfluid to normal phase transition in strongly correlated bosons in two and three dimensions},\ }\href {https://doi.org/10.1103/PhysRevA.86.043629} {\bibfield  {journal} {\bibinfo  {journal} {Phys. Rev. A}\ }\textbf {\bibinfo {volume} {86}},\ \bibinfo {pages} {043629} (\bibinfo {year} {2012})}\BibitemShut {NoStop}%
\bibitem [{\citenamefont {Harada}\ and\ \citenamefont {Kawashima}(1997)}]{PhysRevB.55.R11949}%
  \BibitemOpen
  \bibfield  {author} {\bibinfo {author} {\bibfnamefont {K.}~\bibnamefont {Harada}}\ and\ \bibinfo {author} {\bibfnamefont {N.}~\bibnamefont {Kawashima}},\ }\bibfield  {title} {\bibinfo {title} {{Universal jump in the helicity modulus of the two-dimensional quantum XY model}},\ }\href {https://doi.org/10.1103/PhysRevB.55.R11949} {\bibfield  {journal} {\bibinfo  {journal} {Phys. Rev. B}\ }\textbf {\bibinfo {volume} {55}},\ \bibinfo {pages} {R11949} (\bibinfo {year} {1997})}\BibitemShut {NoStop}%
\bibitem [{\citenamefont {Buchhold}\ \emph {et~al.}(2021)\citenamefont {Buchhold}, \citenamefont {Minoguchi}, \citenamefont {Altland},\ and\ \citenamefont {Diehl}}]{PhysRevX.11.041004}%
  \BibitemOpen
  \bibfield  {author} {\bibinfo {author} {\bibfnamefont {M.}~\bibnamefont {Buchhold}}, \bibinfo {author} {\bibfnamefont {Y.}~\bibnamefont {Minoguchi}}, \bibinfo {author} {\bibfnamefont {A.}~\bibnamefont {Altland}},\ and\ \bibinfo {author} {\bibfnamefont {S.}~\bibnamefont {Diehl}},\ }\bibfield  {title} {\bibinfo {title} {Effective theory for the measurement-induced phase transition of dirac fermions},\ }\href {https://doi.org/10.1103/PhysRevX.11.041004} {\bibfield  {journal} {\bibinfo  {journal} {Phys. Rev. X}\ }\textbf {\bibinfo {volume} {11}},\ \bibinfo {pages} {041004} (\bibinfo {year} {2021})}\BibitemShut {NoStop}%
\bibitem [{\citenamefont {Poboiko}\ \emph {et~al.}(2023)\citenamefont {Poboiko}, \citenamefont {P\"opperl}, \citenamefont {Gornyi},\ and\ \citenamefont {Mirlin}}]{Poboiko_2023}%
  \BibitemOpen
  \bibfield  {author} {\bibinfo {author} {\bibfnamefont {I.}~\bibnamefont {Poboiko}}, \bibinfo {author} {\bibfnamefont {P.}~\bibnamefont {P\"opperl}}, \bibinfo {author} {\bibfnamefont {I.~V.}\ \bibnamefont {Gornyi}},\ and\ \bibinfo {author} {\bibfnamefont {A.~D.}\ \bibnamefont {Mirlin}},\ }\bibfield  {title} {\bibinfo {title} {Theory of free fermions under random projective measurements},\ }\href {https://doi.org/10.1103/PhysRevX.13.041046} {\bibfield  {journal} {\bibinfo  {journal} {Phys. Rev. X}\ }\textbf {\bibinfo {volume} {13}},\ \bibinfo {pages} {041046} (\bibinfo {year} {2023})}\BibitemShut {NoStop}%
\bibitem [{\citenamefont {Poboiko}\ \emph {et~al.}(2024)\citenamefont {Poboiko}, \citenamefont {Gornyi},\ and\ \citenamefont {Mirlin}}]{Poboiko_2024}%
  \BibitemOpen
  \bibfield  {author} {\bibinfo {author} {\bibfnamefont {I.}~\bibnamefont {Poboiko}}, \bibinfo {author} {\bibfnamefont {I.~V.}\ \bibnamefont {Gornyi}},\ and\ \bibinfo {author} {\bibfnamefont {A.~D.}\ \bibnamefont {Mirlin}},\ }\bibfield  {title} {\bibinfo {title} {Measurement-induced phase transition for free fermions above one dimension},\ }\href {https://doi.org/10.1103/PhysRevLett.132.110403} {\bibfield  {journal} {\bibinfo  {journal} {Phys. Rev. Lett.}\ }\textbf {\bibinfo {volume} {132}},\ \bibinfo {pages} {110403} (\bibinfo {year} {2024})}\BibitemShut {NoStop}%
\bibitem [{\citenamefont {Fava}\ \emph {et~al.}(2023)\citenamefont {Fava}, \citenamefont {Piroli}, \citenamefont {Swann}, \citenamefont {Bernard},\ and\ \citenamefont {Nahum}}]{Fava_2023}%
  \BibitemOpen
  \bibfield  {author} {\bibinfo {author} {\bibfnamefont {M.}~\bibnamefont {Fava}}, \bibinfo {author} {\bibfnamefont {L.}~\bibnamefont {Piroli}}, \bibinfo {author} {\bibfnamefont {T.}~\bibnamefont {Swann}}, \bibinfo {author} {\bibfnamefont {D.}~\bibnamefont {Bernard}},\ and\ \bibinfo {author} {\bibfnamefont {A.}~\bibnamefont {Nahum}},\ }\bibfield  {title} {\bibinfo {title} {Nonlinear sigma models for monitored dynamics of free fermions},\ }\href {https://doi.org/10.1103/PhysRevX.13.041045} {\bibfield  {journal} {\bibinfo  {journal} {Phys. Rev. X}\ }\textbf {\bibinfo {volume} {13}},\ \bibinfo {pages} {041045} (\bibinfo {year} {2023})}\BibitemShut {NoStop}%
\bibitem [{\citenamefont {Starchl}\ \emph {et~al.}(2024)\citenamefont {Starchl}, \citenamefont {Fischer},\ and\ \citenamefont {Sieberer}}]{starchl2024}%
  \BibitemOpen
  \bibfield  {author} {\bibinfo {author} {\bibfnamefont {E.}~\bibnamefont {Starchl}}, \bibinfo {author} {\bibfnamefont {M.~H.}\ \bibnamefont {Fischer}},\ and\ \bibinfo {author} {\bibfnamefont {L.~M.}\ \bibnamefont {Sieberer}},\ }\href@noop {} {\bibinfo {title} {Generalized zeno effect and entanglement dynamics induced by fermion counting}} (\bibinfo {year} {2024}),\ \Eprint {https://arxiv.org/abs/2406.07673} {arXiv:2406.07673} \BibitemShut {NoStop}%
\bibitem [{\citenamefont {Kelly}\ \emph {et~al.}(2023)\citenamefont {Kelly}, \citenamefont {Poschinger}, \citenamefont {Schmidt-Kaler}, \citenamefont {Fisher},\ and\ \citenamefont {Marino}}]{kelly2023}%
  \BibitemOpen
  \bibfield  {author} {\bibinfo {author} {\bibfnamefont {S.~P.}\ \bibnamefont {Kelly}}, \bibinfo {author} {\bibfnamefont {U.}~\bibnamefont {Poschinger}}, \bibinfo {author} {\bibfnamefont {F.}~\bibnamefont {Schmidt-Kaler}}, \bibinfo {author} {\bibfnamefont {M.~P.~A.}\ \bibnamefont {Fisher}},\ and\ \bibinfo {author} {\bibfnamefont {J.}~\bibnamefont {Marino}},\ }\bibfield  {title} {\bibinfo {title} {{Coherence requirements for quantum communication from hybrid circuit dynamics}},\ }\href {https://doi.org/10.21468/SciPostPhys.15.6.250} {\bibfield  {journal} {\bibinfo  {journal} {SciPost Phys.}\ }\textbf {\bibinfo {volume} {15}},\ \bibinfo {pages} {250} (\bibinfo {year} {2023})}\BibitemShut {NoStop}%
\end{thebibliography}%
\end{document}